\documentclass{article}
\usepackage{amssymb}
\usepackage{cite,multirow} 
\usepackage[dvips]{graphicx,epsfig}
\usepackage{graphics}
\def\1ad{\mbox{\normalsize $^1$}}
\def\2ad{\mbox{\normalsize $^2$}}
\def\3ad{\mbox{\normalsize $^3$}}
\def\4ad{\mbox{\normalsize $^4$}}
\def\5ad{\mbox{\normalsize $^5$}}
\def\6ad{\mbox{\normalsize $^6$}}
\def\7ad{\mbox{\normalsize $^7$}}
\def\8ad{\mbox{\normalsize $^8$}}

\makeatletter
\@addtoreset{equation}{section}
\makeatother

\newcommand{\ignora}[1]{}
\def\beq{\begin{equation}}                     %
\def\eeq{\end{equation}}                       %
\def\bea{\begin{eqnarray}}                     
\def\eea{\end{eqnarray}}                       
\def\0 {\nonumber}

\begin{document}

\setcounter{page}{0}
\begin{titlepage}
\titlepage
\rightline{hep-th/0607147}
\rightline{Bicocca-FT-06-13}
\rightline{SISSA 42/2006/EP}
\vskip 3cm
\centerline{{ \bf \Large Deformations of conformal theories}}
\vskip 0.4cm
\centerline{{ \bf \Large and non-toric quiver gauge theories}}
\vskip 1cm
\centerline{{\bf Agostino Butti  $^{a}$, Davide Forcella  $^{b}$, Alberto Zaffaroni  $^{a}$}}
\vskip 1truecm
\begin{center}
\em 
 $^{a}$ 
Universit\`{a} di Milano-Bicocca and INFN, sezione di  Milano-Bicocca\\ 
P.zza della Scienza, 3; I-20126 Milano, Italy
\vskip 0.5cm
 $^{b}$ 
International School for Advanced Studies (SISSA / ISAS)\\ 
via Beirut 2, I-34014, Trieste, Italy\\

\vskip .4cm

\vskip 2.5cm
\end{center}
\begin{abstract}
We discuss several examples of non-toric quiver gauge theories
dual to Sasaki-Einstein manifolds with
$U(1)^2$ or $U(1)$ isometry. We give a general method for constructing 
non-toric examples by adding relevant deformations to the toric case.
For all examples, we are able to make a complete comparison between
the prediction for R-charges based on geometry and on quantum field theory.
We also give a general discussion of the spectrum of conformal dimensions
for mesonic and baryonic operators for a generic quiver theory; in the toric 
case we make an explicit comparison between R-charges of mesons and baryons.

\vskip1cm

\end{abstract}
\vskip 0.5\baselineskip

\vfill
 \hrule width 5.cm
\vskip 2.mm
{\small
\noindent agostino.butti@mib.infn.it\\
forcella@sissa.it\\
alberto.zaffaroni@mib.infn.it}
\begin{flushleft}
\end{flushleft}
\end{titlepage}
\large
\section{Introduction}

D3 branes living at the singularity of a Calabi-Yau cone have provided general and interesting results for the AdS/CFT correspondence.
The IR limit of the gauge theory on the world volume of the D3-branes
is dual to type IIB string theory on the near horizon geometry 
$AdS_5\times H$, where $H$ is the Sasaki-Einstein base of the cone \cite{kw,horizon}. The recent growth in the number of explicit examples, with the 
study of the $Y^{p,q}$
and $L^{p,q,r}$ manifolds \cite{gauntlett,CLPP,MSL,benvenuti,kru2,noi,tomorrow},
was accompanied by a deeper general understanding of the correspondence. 
The AdS/CFT correspondence predicts a precise relation between the
central charge $a$, the scaling dimensions of some operators in the
CFT and the volumes of $H$ and of certain submanifolds.  
Checks of this relation have been performed for the known examples of
Sasaki-Einstein metrics 
\cite{benvenuti,bertolini,hananyX,kru,kru2,noi,tomorrow}.
It is by now clear that all these checks can be done without an
explicit knowledge of the metric. a-maximization \cite{intriligator} provides
an efficient tool for computing central and R-charges on the quantum
field theory side. On the other hand, volume minimization \cite{MSY,MSY2} 
provides a geometrical method for extracting volumes from the geometry without 
knowing the metric. 

The cones $C(H)$ admit a $(\mathbb{C}^*)^r$ action with $r=1 \hbox{ or } 2  \hbox{ or } 3 $. We know that there is always at least one  $\mathbb{C}^*$ action.
In all cones over Sasaki-Einstein manifolds, there is in fact 
an isometry corresponding
to the Reeb vector, the geometric dual of the R-symmetry. The Reeb vector
combines with the dilatation to give a  $\mathbb{C}^*$ action. In general 
$r$ can be bigger giving rise to a bigger isometry group $T^r$. 
The case $r=3$ with isometry $T^3$ 
corresponds to the toric case, which is well understood. 
The long standing problem of  finding the correspondence between toric
singularities and quiver gauge theories has been completely solved
using dimer technology \cite{dimers,rhombi,mirror}. The  
brane tilings \cite{dimers} provide an ingenious Hanany-Witten
construction 
of the dual gauge theory. Many invariants, like the number of gauge groups
or multiplicity of fields with given R charge, have simple expression in terms 
of toric data \cite{hananymirror,kru2,aZequiv,rhombi,mirror}.
It is also possible to  provide a general formula for assigning R-charges to the chiral fields of the quiver gauge theory \cite{proc}.  
Moreover, a general proof of the equivalence between a-maximization
and volume minimization  for all toric singularities has been given in \cite{aZequiv}. 
Much less is known about the non toric case. In the case of (non-abelian) 
orbifolds we can find the dual gauge theory by performing a projection, but
for more general non toric singularities not even simple invariants like
the number of gauge groups or chiral fields are known. In this paper we will focus
on various examples  non toric manifolds with isometry $T^2$ and $T^1$ 
obtained by deforming the toric case.

In the first part of the paper we discuss in details the various ways of
comparing the spectrum of conformal dimensions of chiral operators predicted
by the quantum field theory with the information that can be extracted from
supergravity. On the gravity side, we can determine the spectrum of dimensions
of mesons by analyzing the KK spectrum of the compactification on $H$. Alternatively, we can extract the dimensions of baryons by considering D3-branes
wrapped on three cycles in $H$. To study mesons we need to compute the spectrum
of the scalar Laplacian on $H$ while to study baryons we need to compute volumes
of three cycles in $H$. Both methods give a way of determining the 
R-charges of the elementary fields in the gauge theory. The agreement of
the two computations thus gives an intriguing relation between three cycle
volumes and eigenvalues of the Laplacian. In the toric case, where everything
is under control, we will show quite explicitly that this relation is
fulfilled. The proof becomes very simple when all the tools that can be
used in the toric case (the R-charge parametrization in terms of toric data
and the $\Psi$-map for mesons) have been introduced. It would be quite
interesting to understand how the relation between  three cycles
volumes and eigenvalues of the Laplacian can be generalized to the non
toric case, where the understanding of divisors on the cones $C(H)$ is 
still lacking.   

In the second part of this paper we will provide examples of non-toric
 $T^2$ and $T^1$ quiver gauge theories. 
A convenient way of realizing a 
large class of such theories is the following.
We add to a quiver gauge theory dual
to a toric geometry suitable superpotential terms, keeping the same
number of gauge groups and the same quiver diagram. The new terms in
the superpotential must be chosen in such a way to break one or
both of the two $U(1)^2$ flavor symmetries of the original toric
theory: they correspond therefore to a relevant deformation of the
superconformal toric theory which in the IR leads generically to a new
surface of superconformal fixed points, characterized by different
values for the central charge $a$ and for the scaling dimensions of
chiral fields. All conformal gauge theories with a three 
dimensional moduli space of vacua and with a supergravity dual are described
by Calabi-Yau and thus the dual background is AdS$_5\times H$ with
$H$ Sasaki-Einstein \footnote{We thank A. Tomasiello for
an enlightening discussion on this point. The supergravity description of the new quiver gauge theory can be realized
in terms of a new Sasaki-Einstein manifold $H$, which is our interest in this
paper, or more generally in terms
of warped AdS$_5$ backgrounds with non vanishing three form fluxes.
Generic relevant deformations, which reduce the dimension
of the moduli space of vacua of the gauge theory, are 
typically described by an AdS$_5$
background with three-form flux. Conformal theories with three dimensional
moduli space are instead necessarily of Calabi-Yau type.
This 
observation is based on the analysis of the supersymmetry conditions
for supergravity solution with fluxes
\cite{tommi,martucci,susyads5,su2}. See Section 4 for more
details.}. Our strategy for producing new examples of 
non toric quiver gauge theories will be as following: in the family of
relevant deformations of a toric case that lead to IR fixed points we will
select those cases where the moduli space is three dimensional. We will
provide examples based on delPezzo cones, Generalized Conifolds, $Y^{p,q}$
and $L^{p,q,r}$ manifolds. We will also show that this procedure is quite
general: given a large class of toric quiver theories we can choose
relevant deformations that lead to examples of non toric manifolds
with $T^2$ or $T^1$ 
isometry. The dual gauge theory has the same quiver than the original theory
but differs in the superpotential terms \footnote{Obviously, this
  means that we are not considering the most general non toric
  theory.}.  
We will give a complete
characterization of this
particular class of $T^2$ and $T^1$ theories and we will compute central and R-charges and characters
by adapting toric methods. The Calabi-Yau corresponding to the new 
IR fixed points can be written as a system of algebraic equations using  
mesonic variables. We in general obtain a non complete intersection variety. 
We can confirm that the manifold is indeed a Calabi-Yau by comparing the
volume extracted from the character (under the assumption of dealing with
a Sasaki-Einstein base) with the result of a-maximization.

There are various differences between theories with different number of 
isometries which we will encounter in our analysis.
These differences are particularly manifest in
spectrum of chiral mesons. The chiral ring can be mapped 
to the cone of holomorphic functions on $C(H)$. The partition
function (or character) counting chiral mesons 
with given $U(1)^r$ charge contains several information about the dual gauge
theory. In particular, as shown in \cite{MSY2}, the volume of $H$ as a 
function of the Reeb vector can be extracted from the character. We can 
then perform volume minimization, whose details depend on the value of $r$. 
The minimization is done on two parameters in the toric case
but it is done on zero parameters - so it is not even a minimization
anymore - in the case of $T^1$.
As a result the central charge of $T^1$ cases is always rational \footnote{This statement has a counterpart
in a-maximization. In fact one can eliminate the baryonic symmetries from the
process of a-maximization  using linear equations \cite{aZequiv}. 
The number of free parameters is then equal to the
number of flavor symmetries which is $r-1$, reducing to zero for $r=1$.}.
The spectrum of dimensions of the chiral mesons of the dual gauge theory is also sensitive to the number of 
isometries. In fact, there is a  simple formula 
$$\Delta=\sum_{i=1}^r b_i m_i$$ 
that expresses the dimension of a meson in terms of its integer charges $m_i$ 
under the $U(1)^r$ torus and the components of the Reeb
vector $b_i$ in the same basis. In the toric case, the cone of
holomorphic functions is given by the dual of the fan,  ${\cal C}^*$,
and there is exactly one holomorphic function for each point in
$\cal{C}^*$ \cite{fulton}. The character can be evaluated as discussed
in \cite{MSY2}. In the $T^1$ case, 
instead, it follows from the relation between dimension and Reeb that
all the dimensions of mesons are integer multiples of a given
quantity. The cone 
of holomorphic functions in this case is just an half-line, with multiplicities
typically greater than one.
The intermediate case  $r=2$ has some interesting features since the cone of
holomorphic functions has dimension greater than one and there is at least
one flavor symmetry requiring Z-minimization. 

The paper is organized as follows. In Section 2 we review the
description of mesons in terms of KK modes and holomorphic
functions. We also derive a useful 
formula relating the dimensions of mesons to the Reeb vector. 
In Section 3 we compare, in the toric case, the spectrum of conformal
dimensions of chiral operators that can be extracted from mesons and
from baryons.  
In Section 4 we describe the general method for obtaining non toric theories
with $T^2$ or $T^1$ isometry by deforming toric ones. In Section 5 we discuss
in all details the case of quivers associated to cones over the delPezzo
manifold $dP_4$, the blow-up of $\mathbb{P}_2$ at four points. When the four
points are in specific positions we obtain the toric model $PdP_4$.
This can be deformed both to a $T^2$ and a $T^1$ theory by changing
the superpotential \footnote{The $T^1$ theory obtained in this way is
  the usual cone over $dP_4$, 
the blow-up op $\mathbb{P}_2$ at four generic points, whose dual gauge theory
was already known \cite{wijnholt}.}. 
In Section 6 and 7 we provide other examples
based on Generalized Conifolds, $Y^{p,q}$ and $L^{p,q,r}$ manifolds and we
will argue that this procedure for obtaining non toric examples is quite
general. In Section 8 we generalized our proof of the equivalence between
a-maximization and Z-minimization to these more general theories. Finally,
in Appendix A we review the basic facts about the toric case and in appendix B 
we collect some technical details of the moduli space analysis.

\section{The spectrum of chiral mesons}
\label{mesons}

In this Section we discuss the dual description of mesons of a quiver
gauge theory in terms of holomorphic functions on the Calabi-Yau and
we give useful formulae for their dimensions. We review this
description in the  
following, collecting various arguments that have appeared elsewhere in 
the literature. A similar point of view was taken in the very recent \cite{GMSY}.

In all the paper we consider $N$ D3-branes living at the tip of a CY cone $C(H)$.
The base of the cone, or horizon, is a five-dimensional
compact Sasaki-Einstein manifold $H$ \cite{kw,horizon}.
The IR limit of the gauge theory living on the branes is 
${\cal N}=1$ superconformal and dual in the AdS/CFT correspondence 
to the type IIB background $AdS_5\times H$. 
For all the cones $C(H)$ the gauge theory is of quiver type, with bifundamental matter fields.


The connection of the mesons to holomorphic functions is easily understood
from the quantum field theory point of view. The superconformal gauge theory
living on a D3 brane at conical singularities has a moduli
space of vacua which corresponds to the arbitrary position of the 
brane on the cone; it is  isomorphic to the $CY$ cone. 
It is well known that the moduli space of vacua of an $\mathcal{N}=1$ gauge 
theory can be parameterized by a complete set of operators invariant under
the complexified gauge group. The moduli space of vacua is then
described by the chiral mesonic operators \footnote{The non-abelian theory
of N D3 branes has obviously a larger moduli space. In addition to the
mesonic vacua corresponding to N copies of the cone, we also have baryonic
flat directions. For the abelian theory on a D3 brane, baryonic operators
are vanishing and the moduli space is described by the mesons.}. 
These can be considered as well defined functions on the cone: the
functions assign to every point in the moduli space the v.e.v. of the
mesonic operators in that vacuum. From the supersymmetry of the vacuum
it follows that these functions are the same for $F$-term equivalent
operators. We thus expect a one to one correspondence between the
chiral ring of mesonic operators (mesonic operators that are
equivalent up to $F$-terms equations) and the holomorphic functions on
the cone.  

When we restrict the holomorphic function to $H$ we obtain an eigenvector
of the Laplacian $\Box_{H}$. The explicit relation is as follows.
Because the cone is K\"ahler the Laplacian  on $C(H)$ is 
$$\Box_{C(H)}= \bar{\partial } \bar{\partial}^+ + \bar{\partial}^+ \bar{\partial }$$
 and every holomorphic function $f(z)$ on the cone is an harmonic function: 
$$\Box_{C(H)} f(z) = 0\, .$$ 
By writing the $CY$ metric on the cone as $ds^2_{C(H)}= dr^2 + r^2ds^2_H $
we can explicitly compute
\begin{equation}\label{lap6}
\Box_{C(H)} f(z)=\Big( \frac{\partial ^2}{\partial r ^2} + \frac{5}{r}\frac{\partial}{\partial r} + \frac{1}{r^2} \Box_{H} \Big) f(z) = 0\, . 
\end{equation}   
We can rewrite the previous equation as follows
\begin{equation}\label{lap56}
\frac{1}{r^2} \Big( \xi ( \xi + 4 ) + \Box _{H} \Big) f(z) = 0 
\end{equation}
where $\xi = r\partial _{r}$ is the dilatation vector field.
An holomorphic function with given scaling dependence
$$\xi f(z) = \delta f(z)\, ,$$
when restricted to the base $H$, becomes 
an eigenvector of the Laplacian operator
\begin{equation}\label{lap5}
\Box _{H} f(z) | _H = - E f(z) | _H  
\end{equation}   
with eigenvalue $ E = \delta ( \delta + 4 )$. We are now going to show that
$\delta$ is the conformal dimension of the corresponding meson.

We first identify the restriction of the
holomorphic function to the base with the KK harmonic corresponding to the
meson according to the AdS/CFT correspondence. In the gravity side, in
fact, we have the type IIB string theory propagating on the background
$AdS_5 \times H$, and at low energy we can consider the supergravity
regime and compute the $AdS$-spectrum making the $KK$ reduction on the
compact space $H$.  
Every ten dimensional field is decomposed in a complete
basis of eigenvectors of the differential operator on $H$ corresponding to
its linearized equation of motion \footnote{After a suitable
  diagonalization, of course.}. For scalar fields, this operator is
just related to the Laplacian.  
From the eigenvalue of the differential operator we read the mass $m$
of a scalar in $AdS_5$ which is related to the dimension $\Delta$ of the
dual field theory operator by the familiar 
formula $m^2 = \Delta ( \Delta - 4)$. 

Consider an eigenvector of the Laplacian $Y_I$ with eigenvalue $E$ 
$$\Box_H Y_I = - E\, Y_I\, .$$
In the same way as in the familiar case $H=S^5$ the k-th harmonic on $S^5$
is mapped to the operator $\phi_{\{i_1}...\phi_{i_k\}}$,
the function $Y_I$ can be mapped to particular operators $O_I$ made with
the scalar fields of the gauge theory. The same eigenvector $Y_I$ 
appears in the KK decomposition of various ten dimensional fields.
In particular, it appears in the expansion of the dilaton and also of  
the fields $g_\alpha^\alpha$ and
$F^{(5)}_{\alpha\beta\gamma\delta\epsilon}$.  From the general
analysis in \cite{kim,gubs,ferrara}, we have the following
correspondence between fields, dual operators 
and dimensions \footnote{The result in the table can be understood as follows.
The expansion of the dilaton is simply given in terms of eigenvalues
of the internal Laplacian $\Box_H$. The fields $g_\alpha^\alpha$ and 
$F^{(5)}_{\alpha\beta\gamma\delta\epsilon}$ instead mix in the equations of 
motion, which must be diagonalized and produce two KK towers. The 
five-dimensional mass is still related to the internal Laplacian with 
corrections due to the curvature of the internal manifold. 
The difference in dimensions of the various towers are compatible with the
fact that, if we assign dimension $\delta$ to $O_I$ then $F_{\mu\nu}^2 O_I$
has dimension $\delta+4$ and $F_{\mu\nu}^4 O_I$
has dimension $\delta+8$. 
Finally, 
the identification of scalars fields and operators is suggested by the 
couplings in the
Born-Infeld action to supergravity modes. The scalar fields on a D3 brane probe
identifies its position in the internal manifold and operators made with
scalar fields can be identified with functions on $H$.}
\vskip0.5truecm
\begin{tabular}{|c|c|l|l|l|}
\hline 
$\phi$ & $F_{\mu\nu}^2 O_I$ &  $m^2=E$ &  $E=\Delta (\Delta-4)$ & $\Delta= \delta+4$ \\[1em]
$g_\alpha^\alpha/F^{(5)}$  & $O_I$ & $m^2= 16 + E \mp 8 \sqrt{4+E}$ & $E  = \Delta (\Delta+4)$ & $\Delta=\delta$\\
 \, & $F_{\mu\nu}^4 O_I$ & \, & $E  =  (\Delta-4)(\Delta-8)$ & $\Delta=\delta+8$\\
\hline
\end{tabular}
\vskip0.5truecm
The case of chiral mesons $M_I$ corresponds to the 
subset of eigenvectors $Y_I$ that are restriction of holomorphic functions 
on $C(H)$. We see from the previous table that the dimension $\Delta$ of
a meson is related to the corresponding Laplacian eigenvalue by
$E=\Delta(\Delta+4)$. Comparison with equation (\ref{lap5}) shows that
the scaling behavior  
$\delta$ of an holomorphic function coincides with the conformal 
dimension of the corresponding meson: $\Delta\equiv \delta$.

It is important to observe that the conformal dimension of the meson can be 
fixed in terms of the its charge quantum numbers. We are assuming that
there exists a $(\mathbb{C}^*)^r$ action with $r=1 \hbox{ or } 2
\hbox{ or } 3 $ on $C(H)$. We know that there is at least one
$\mathbb{C}^*$ action which pairs the 
dilatation and the Reeb vector of the Sasaki-Einstein manifold. The Reeb
vector $K$ is the geometric dual of the R-symmetry of the gauge theory. 
In case of extra $U(1)$ 
isometries, $r$ can be bigger, $r=3$ corresponding to the toric case.   
If we define $\phi _k $ with $k=1,...,r$ the coordinates relative to
the action of $T^r\subset (\mathbb{C}^*)^r$, a function $f(z)$ with
charges $m_k$  
satisfies \begin{equation}\label{caricf}
\partial _{\phi _k} f(z) = i m_k f(z)
\end{equation}   
In these coordinates we can decompose the Reeb vector $K$ as $K = \sum
_{k=1} ^r b_k \partial _{\phi _k }$ \cite{MSY2}. Using this
decomposition we have: 
\begin{equation}
K f(z) = i \sum _{k=1} ^r b_k m_k f(z) = i (m,b)f(z) 
\end{equation}   
\\
For a Sasaki-Einstein cone 
\begin{equation}\label{JK}
\xi = - J K  
\end{equation} 
where $J$ is the complex structure of the $CY$ three-fold \cite{horizon,MSY}. We thus
 obtain the following important relation \footnote{A generic vector
 field over the cone $V$ can be written as $V=\sum _{i=1}^3 v_i
 \partial _{z_i}  + \sum _{i=1}^3 \bar{v}_i \partial _{\bar{z}_i}$
 where $v_i \hbox{, } \bar{v}_i$ are some coefficients. By definition
 we have $J V = i \sum _{i=1}^3 v_i \partial _{z_i } -i \sum _{i=1}^3
 \bar{v}_i \partial _{\bar{z}_i }$, and if we apply this differential
 operator on any holomorphic function $f(z)$ we obtain $J V f(z) = i V
 f(z)$.}:
\begin{equation}\label{reebm}
\xi f(z) = - J K f(z) = (m,b) f(z) 
\end{equation} 
This imply that the dimension $\Delta $ of any meson with charges $m_k$ 
satisfies the relation $$\Delta = (m,b)$$ or
equivalently for the R charge 
\begin{equation}\label{R}
 R = \frac{2}{3} \hbox{ }\Delta = \frac{2}{3} \hbox{ }(m,b) 
\end{equation}   
Using this relation it is possible to compute the exact $R$-charges
for every $BPS$ mesonic operators of a gauge theory dual to an
arbitrary $CY$ conical singularity as a function of the Reeb vector
$\vec{b}$ of the CY metric. On the other hand, the Reeb vector can be  
computed from the geometry resolving an extremization problem
explained in detail in \cite{MSY2}: the volume of a Sasaki metric over
$H$ is a function only of the Reeb vector $\mbox{Vol}_H(\vec{b})$, and
the Reeb vector corresponding to a Sasaki-Einstein metric on $H$ is
the minimum of  $\mbox{Vol}_H(\vec{b})$ when $\vec{b}$ varies on a
suitable $r-1$ dimensional affine subspace of $\mathbb{R}^r$. Moreover,  
the function $\mbox{Vol}_H(\vec{b})$ can be computed without knowing the 
metric \cite{MSY2}.  

For singularities with only an $U(1)$ action formula (\ref{R}) implies
that the mesonic $R$-charges are all integer  
multiples of a common factor. This fact can be easily tested in the
known cases i.e. the complex cones over the non-toric del Pezzo
surfaces \cite{fracohidexcept}. In the toric case (\ref{R}) is a known
result and it was explained in \cite{Heun eq} in the case of the
$L^{p,q,r}$ singularity. 
In this paper we will use this result also in the more interesting
case of the varieties that admit a $T^2$ action. 

\section{Baryons and mesons: the toric case}
\label{barmes}

We have seen in the previous Section that AdS/CFT predicts an equality
(\ref{R}) between the scaling dimensions $\Delta$ of mesons in the gauge
theory and some geometrical quantities ($\vec{b} \cdot \vec{m}$) related to the
eigenvalues of the scalar Laplacian on the Sasaki-Einstein $H$. In the
toric case, where the AdS/CFT correspondence has been explicitly
built, equation (\ref{R}) can be proved, as we show in this Section, thus checking the matching of
mesonic operators in field theory with suitable supergravity states in
string theory (the check of the matching between baryonic
operators and wrapped D3-branes inside $H$ was performed in
\cite{aZequiv} for the toric case).
These results also suggest a quite unexpected and suggestive connection between Laplacian
eigenvalues and volumes of cycles in a Sasaki-Einstein manifold.

In the toric case the geometry can be described by the
fan $\mathcal{C}$, a polyhedral cone in $\mathbb R^3$, generated by the integer vectors
$\vec{V}_i=(x_i,y_i,1)$, $i=1,\ldots d$. The toric diagram $P$ is the
convex polygon with vertices $(x_i,y_i)$. The dual cone $\mathcal{C}^*$ is the image of the
momentum map and its integer points $\vec{m}\in \mathcal{C}^*$ are the charges
of holomorphic functions over $C(H)$: it is known from toric geometry that holomorphic functions on $C(H)$ are in 
one-to-one correspondence with points of $\mathcal{C}^*$
\cite{fulton}, and the multiplicity of each point is equal to one. See Appendix \ref{dimtor} for more details about
these geometrical tools and for a review of periodic quivers and
dimers we will use in the following.

Before proving equation (\ref{R}), we have to
describe explicitly the correspondence, introduced in the previous Section, between holomorphic functions
on $C(H)$ and mesons in field theory. This
correspondence is known in the literature and is called $\Psi$-map
\cite{hanherve,ago} \footnote{The $\Psi$-map was originally defined in
  \cite{hanherve} as a linear function that
  assigns to every path in the quiver a divisor in the geometry.  
We will use the equivalent definition in
\cite{ago} that assigns to paths in the quiver their trial charge.}.  
Recall that mesons in the gauge theory are closed
oriented loops in the periodic quiver  
drawn on $T^2$. To each link in the periodic quiver we can
assign a trial charge written as a sum of suitable parameters $a_i$,
$i=1,\ldots d$ in one to one correspondence with the edges $V_i$ of
$\mathcal{C}$: with the restriction $\sum a_i=2$ this is a
parametrization of R-charges. If we compute the trial R-charge $\Psi(M)$
of a meson $M$ we discover that it can be written as \cite{ago}: 
\begin{equation}\label{psi}
\Psi(M) =\sum_{i=1}^d (\vec{m},\vec{V}_i) a_i 
\end{equation} 
for a point $\vec{m}=(n,m,c)\in \mathcal{C}^*$, where $(n,m)$ are the homotopy numbers of the
meson on the torus $T^2$ of the periodic quiver. We have therefore found the map from mesons
to integer points in $\mathcal{C}^*$, that is holomorphic functions on
$C(H)$. Moreover F-term equivalent
mesons are mapped to the same point $\vec{m}$ and conversely mesons
mapped to the same point are F-term equivalent, so that there is a one
to one correspondence between the chiral ring of mesons in the gauge
theory and the semi group of integer points of $\mathcal{C}^*$.

Using the $\Psi$-map to compute the charges, equation (\ref{R})
becomes:
\begin{equation}
\sum_{i=1}^d (\vec{m},\vec{V}_i) \bar a_i = \frac{2}{3} \vec{m} \cdot
\vec{b}
\label{matex}
\end{equation}
To compare the two sides of this equation we need to compute 
the exact charges $\bar a_i$ from a-maximization \cite{intriligator}, and 
the vector $\vec{b}$ from volume minimization \cite{MSY}.

\begin{figure}
\begin{center}
\includegraphics[scale=0.6]{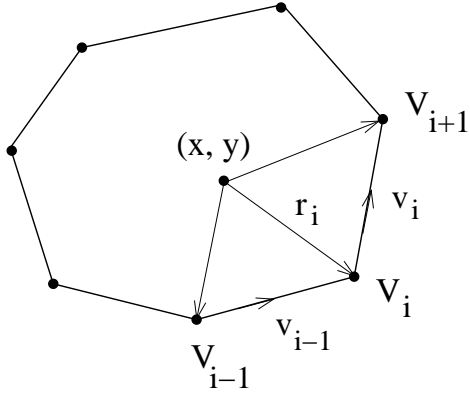} 
\caption{The toric diagram $P$ is a convex polygon in the plane with
  integer vertices obtained as the intersection of the toric fan
  $\mathcal{C}$ with the plane $z=1$.}
\label{toricd}
\end{center}
\end{figure}

As already mentioned, the Reeb vector for a Sasaki-Einstein metric 
can be found by minimizing the volume function $\mbox{Vol}_H(\vec{b})$
of the Sasaki manifold $H$ on
a suitable affine subspace of $\mathbb{R}^r$. 
In the case of toric geometry, volume minimization, also known as
Z-minimization, is performed by varying $\vec{b}=3(x,y,1)$ with
$(x,y)$ inside  
the toric diagram $P$ \cite{MSY}. For convenience,
let us introduce the sides of the toric
diagram: $v_i\equiv(x_{i+1},y_{i+1})-(x_i,y_i)$ and the vectors $r_i$
joining the trial Reeb $(x,y)$ with the vertices:
$r_i\equiv(x_i,y_i)-(x,y)$, 
see Figure \ref{toricd}; we will use the notation: $\langle v,w
\rangle \equiv \det (v,w)$ for vectors in the
plane of the toric diagram $P$. 

a-maximization is done instead on $d-1$
independent trial R-charges $a_i$, but as shown in \cite{aZequiv}, it
can be consistently restricted to a two dimensional space of
parameters. The maximum of the
central charge $a(a_i)$ as function of the trial charges $a_i$ lies on the surface parameterized as 
\begin{equation}
a_i \rightarrow a_i(x,y)\equiv \frac{2 l_i}{\displaystyle \sum_{j=1}^d
  l_j} \qquad \textrm{with} \qquad l_i\equiv\frac{\langle v_{i-1},
  v_{i} \rangle }{ \langle r_{i-1},v_{i-1} \rangle \langle r_i, v_i
  \rangle} 
\label{aixy1}
\end{equation} 
Moreover  for every $(x,y)$ inside the toric diagram $P$ we have\cite{aZequiv}:
\begin{equation}
a(x,y)=\frac{\pi^3}{4 \mbox{Vol}_H(x,y)} \qquad \textrm{with} \qquad a(x,y)\equiv a(a_1(x,y),\ldots a_d (x,y))
\label{matcht31}
\end{equation}
where $\mbox{Vol}_H(x,y)$ is the volume function
$\mbox{Vol}_H(\vec{b})$ for a Sasaki metric computed at the trial Reeb
vector: $\vec{b}=3(x,y,1)$ and $a(x,y)$ is the 
central charge $a$ evaluated on the parameterization $a_i(x,y)$. This shows
the equivalence of a-maximization and Z-minimization for all toric singularities \cite{aZequiv}.

Now it is easy to see that equation (\ref{matex}) holds not only at
the extremal points $\bar{a}_i$ of a-maximization and $(\bar x,\bar y)$ of Z-minimization, but on the entire
two dimensional surface parameterized by $(x,y)\in P$. Indeed (\ref{matex}) follows easily from the equality:
\begin{equation}
\sum_{i=1}^d a_i(x,y) \vec{V}_i = 2 (x,y,1) \qquad \forall (x,y) \in P
\end{equation}
which is easily proved \cite{MSY,aZequiv}: the third component is
obvious and the first two reduce to the geometrical equality: $\sum_i
l_i r_i=0$, which is equation (A.27) in \cite{aZequiv}. We have therefore checked the matching of conformal dimensions of mesons 
with the masses of corresponding supergravity states. 

In the case of toric geometry, one could also extract information about
the exact R-charges of the chiral fields in the quiver gauge theory by 
studying baryonic operators. It is known indeed that baryons
are dual to D3-branes wrapped over calibrated three cycles
$\Sigma \subseteq H$. 
In the toric case there is a basis of calibrated cycles
$\Sigma_i$, $i=1,\ldots d$ which are in one to one correspondence with vertices
$\vec{V}_i$ of the toric diagram. The baryonic
operator in field theory: $\epsilon^{\alpha_1...\alpha_N}
\Phi_{\alpha_1}^{\beta_1}...\Phi_{\alpha_N}^{\beta_N} 
\epsilon_{\beta_1...\beta_N}$ made with a chiral bifundamental field
$\Phi_\alpha^\beta$ with trial charge $a_i$ is dual to a D3-brane wrapped on 
$\Sigma_i$ \footnote{If the baryon is
built with a chiral field of trial charge $a_i+a_{i+1}+\ldots a_j$, it
corresponds to a D3-brane wrapped over the union $\Sigma_i \cup
\Sigma_{i+1} \cup \ldots \Sigma_j$.}. The AdS/CFT relation between the
scaling dimension of the operator and the mass of the dual state is
\cite{klgub}:
\begin{equation}
a_i= \frac{\pi}{3} \, \frac{\hbox{Vol}(\Sigma_i)}{\hbox{Vol}(H)}
\label{bardim}
\end{equation}   
This relation is perfectly consistent with the parameterization (\ref{aixy1})
since, as shown in \cite{MSY}, the formulae for the
volumes of $H$ and of $\Sigma_i$ as a function of the Reeb vector $3(x,y,1)$
are:
 \begin{equation}
\hbox{Vol}_{\Sigma_i}(x,y)= \frac{2 \pi^2}{9} l_i \qquad
\hbox{Vol}_{H}(x,y)=\frac{\pi}{6}\sum_{i=1}^d \hbox{Vol}_{\Sigma_i} (x,y)
\label{volfun}
\end{equation}
Therefore, the relation (\ref{bardim}) is valid, not only at the extremal point $(\bar x, \bar y)$, but also
for generic points $(x,y)\in P$.

As we have just seen the exact charges of chiral fields in the
quiver gauge theory, which allow to
compute scaling dimensions of baryons and mesons, are related to
volumes of calibrated submanifolds (\ref{bardim}) inside $H$ (baryons)
and to (certain) eigenvalues of the scalar Laplacian on $H$ (mesons). We can
invert these relations and get an interesting relation only between
the geometrical quantities. 

We use points $\vec{m}$ in $\mathcal{C}^*$ to parametrize mesons. 
From the previous Section we have the relation between some eigenvalues
$-E_{\vec{m}}$ of the Laplacian on $H$ and the $R$-charges $R_{\vec{m}}$ of the
corresponding mesonic operators:  
\begin{equation}\label{eigvrc}
E_{\vec{m}}= \Delta (\Delta + 4) = \left( \frac{3}{2}R_{\vec{m}}  + 2 \right)^2 -4
\end{equation}   
where we have used $\Delta=3/2 R_{\vec{m}}$. The R-charge $R_{\vec{m}}$
for mesons is computed through the $\Psi$-map (\ref{psi}), and using
equation (\ref{bardim}) for the $a_i$ we get the relation:
\begin{equation}\label{masterrel}
E_{\vec{m}}= \Big[ \Big( \sum_{i=1}^d (\vec{m},\vec{V}_i)\frac{\pi
  }{2} \frac{ \hbox{Vol}(\Sigma _i )}{\hbox{Vol}(H)} +2 \Big) ^2 - 4
  \Big]  
\end{equation}   
This is a very interesting general geometric relation that is given by
the $AdS/CFT$ correspondence; we have seen that it is valid for every
six dimensional toric $CY$ cone.

We can try to invert equation (\ref{masterrel}) in order to find
information concerning the divisors and their volumes starting from
the eigenvalues $E_{\vec{m}}$. This is a very interesting point of
view, because in the toric case we can directly compute both sides of
the relation without the explicit knowledge of the Sasaki-Einstein
metric on the base $H$, but in the general case we are able to compute
only the left hand side. 
Therefore it would be interesting to try to understand
if it is possible to generalize equation (\ref{masterrel}) outside the
toric geometry and find in this way a tool to obtain information
regarding the divisors in the general case of a $CY$ conical
singularity without knowing the explicit $CY$ metric.   

More generally the problem of computing the volumes of the bases of
divisors for non toric cones without the explicit knowledge of the
$CY$ metric should be further investigated. When the Sasaki-Einstein manifold
$H$ is quasi-regular (which includes all $T^1$ theories) it can be
realized as a $U(1)_R$ fibration over a K\"ahler-Einstein orbifold basis,
and the computation of volumes can be solved through an
intersection problem on this basis, as shown in
\cite{Herzog1,Herzog2,intriwecht}. For instance in \cite{Herzog1} one can find
the computations of volumes of divisors for the generalized conifolds, that we
will consider in Section \ref{lop}. For irregular $T^2$ cases it would be
interesting to know whether there exist localization methods for the
volumes of divisors analogous to those in \cite{MSY2} for
$\textrm{Vol}(H)$.

\section{Non-toric cones}
\label{mod}
In this Section we explain how to build a wide class of non-toric
examples of the AdS/CFT correspondence starting from toric cases. 
The idea is very simple: we have to add to a quiver gauge theory dual
to a toric geometry suitable superpotential terms, keeping the same
number of gauge groups and the same quiver diagram (some fields
may be integrated out if we are adding mass terms). The new terms in
the superpotential must be chosen in such a way to break one or
both of the two $U(1)^2_F$ flavor symmetries of the original toric
theory: they correspond therefore to a relevant deformation of the
superconformal toric theory which in the IR leads generically to a new
surface of superconformal fixed points, characterized by different
values for the central charge $a$ and for the scaling dimensions of
chiral fields. 

We will be interested in particular to the cases when the mesons of
the new theory obtained by adding superpotential terms still describe
D3-branes moving in a complex three dimensional cone.
To compute this
geometry in concrete examples it is easier to replace all $SU(N)$
gauge groups with $U(1)$ gauge groups and compute the (classical)
moduli space of the quiver gauge theory; this is the geometry seen by
a single probe D3-brane. Moreover in the abelian case the baryonic
operators are automatically excluded. 

With a generic superpotential the complex
dimension of this mesonic moduli space can be less than three. This 
happens for example with massive deformations. Moreover, inside
 a manifold of fixed points, in addition to gauge theories with three 
dimensional moduli space there are other theories with reduced space of vacua.
In fact, the addition of relevant or marginal deformations to a gauge theory
alters the F-term equations and this easily leads to a reduced or even to 
a zero dimensional moduli space. 
Familiar cases studied in the literature are the
mass deformation of $\mathcal N=4$ theory \cite{kpw} and the $\beta$ 
deformation of $\mathcal N=4$ theory or of other toric cases 
\cite{maldalunin,conformal}. In all these cases, the supergravity dual 
is a warped $AdS_5$ solution with three-form flux; the 
fluxes in the internal geometry generate 
a potential for the probe D3-brane that cannot move in
the whole geometry. 
Different is the story when the moduli space of the
gauge theory is three dimensional: the supergravity dual of a superconformal
gauge theory with three dimensional moduli space is necessarily 
constructed using a Calabi-Yau cone \footnote{We thank A. Tomasiello for
an enlightening discussion on this point. The argument goes roughly as follows.
Supersymmetric solutions of type IIB can be described by SU(2) or SU(3) 
structures and a very convenient framework to describe them is given by 
the pure spinor formalism \cite{tommi}. Supersymmetric conditions for D3
brane probes in general backgrounds have been discussed in \cite{martucci}.
It is easy to check (see for example \cite{su2}) that all SU(2) structure
solutions have D3-brane moduli space of dimension less than three; massive
and $\beta$-deformations are indeed of SU(2) type \cite{su2}. It follows
that all solutions with three dimensional moduli space for D3 probes
are necessarily SU(3) structures. However,  
it has been proven in \cite{susyads5} that all $AdS_5$ solutions
with SU(3) structure (this are SU(2) structure in the five-dimensional language
used in \cite{susyads5}) are necessarily of the form $AdS_5\times H$ with
$H$ Sasaki-Einstein. This proves the argument.
Notice that the requirement of conformal invariance is
necessary for this argument; there are well known examples of 
non-conformal SU(3) structure solutions which are not of Calabi-Yau type, 
for example the Maldacena-Nunez solution \cite{MN} or the
baryonic branch of the Klebanov-Strassler solution \cite{baryonic}.}. 
The gauge theory is therefore dual
to type IIB string theory on $AdS_5 \times H$ (with no $H_3$ or $B_3$
fluxes and constant dilaton) where $H$ is the Sasaki-Einstein horizon
of the CY cone; the isometry group of $H$ is now reduced to $U(1)^2$
or $U(1)$ if we have broken one or two flavor symmetries
respectively. The CY cone is therefore no more toric.

In the construction of new examples of non-toric quiver gauge theories with
this method we obviously need to be careful about the real existence 
of the IR fixed point. We can compute central and R-charge of the
IR theory by using a-maximization: we have to check that all 
unitarity requirements are satisfied. Obstructions to the existence
of CY conical metrics on singularities, which are the geometric duals 
of the unitarity constraints, have been discussed in \cite{GMSY}. 
 
We explain now the details of the construction.
Consider the original toric theory. As explained in the previous Section, 
the mesons of this theory are closed oriented loops in the periodic quiver
drawn on $T^2$ and are mapped by the $\Psi$-map into integer points
$\vec{m}$  of the polyhedral cone $\mathcal{C}^*$, the dual cone of the toric
fan $\mathcal{C}$. The trial charge of any meson mapped to $\vec{m}$ is 
\begin{equation}
\Psi = \sum_{i=1}^d \left( \vec{m} \cdot \vec{V}_i \right) a_i
\label{psimap}
\end{equation}
expressed in terms of the parameters $a_i$, $i=1, \ldots d$ associated
with the vertices of the toric diagram $P$ as in \cite{aZequiv}; $d$ is
the number of vertices of $P$ and the vectors $\vec{V}_i=(x_i,y_i,1)$ are
the generators of the fan $\mathcal{C}$.

The mesons belonging to the superpotential of the toric theory are
those mapped to the point $\vec{m}_0 \equiv (0,0,1)$, since their
trial charge is $a_1+\ldots a_d$ \cite{hanherve,ago}. 
Therefore we have to impose the conditions:
\begin{equation}
\sum_{i=1}^d \left( \vec{m}_0 \cdot \vec{V}_i \right) a_i =
\sum_{i=1}^d a_i=2 \qquad \textrm{R-charges}
\label{erre}
\end{equation}  
to find the non anomalous R-symmetries or
\begin{equation}
\sum_{i=1}^d \left( \vec{m}_0 \cdot \vec{V}_i \right) a_i =
\sum_{i=1}^d a_i=0 \qquad \textrm{global charges}
\label{global}
\end{equation}
to parametrize the $d-1$ global non anomalous $U(1)$ symmetries. Among
them, the $d-3$ baryonic symmetries satisfy the further constraint:
\begin{equation}
\sum_{i=1}^d a_i \vec{V}_i=0  \qquad \textrm{global baryonic charges}
\label{baryonic}
\end{equation}

Suppose now to add to the superpotential (all) the mesons that are
mapped to a new integer point $\vec{m}_1$ of $\mathcal{C}^*$ \footnote{We are
  changing here the theory, but we consider the $\Psi$-map of the
  original toric theory.}. Call $\vec{d}$ the difference:
$\vec{d}=\vec{m}_1-\vec{m}_0$; for the new superpotential terms we
have to impose that the trial charge (\ref{psimap}): $\Psi= \sum
\left( \vec{m}_1 \cdot \vec{V}_i \right) a_i$ is equal to $2$ or to $0$ if
we want to parametrize R-symmetries or global symmetries respectively.
Taking the difference with equations (\ref{erre}) or (\ref{global})
respectively we obtain that the new condition we have to impose to
move away from the toric case is:
\begin{equation}
\sum_{i=1}^d \left( \vec{d} \cdot \vec{V}_i \right) a_i=0
\label{new}
\end{equation}
both for R-symmetries and for global charges. Note that, with the only
constraint (\ref{new}), we can add to the superpotential all mesons
mapped to the points $\vec{m}_k=\vec{m}_0+k \, \vec{d}$ with $k$ integer,
that is all integer points in $\mathcal{C}^*$ lying on a line passing through
$\vec{m}_0$.

Since we are not changing the quiver, the conditions for a charge to
be non anomalous are the same as in the toric case and so are again
satisfied with the known parametrization with $a_i$ satisfying conditions
(\ref{erre}) or (\ref{global}). Note moreover that all the $d-3$
baryonic symmetries (\ref{baryonic}) of the toric case satisfy also
the new restriction (\ref{new}). Therefore imposing condition
(\ref{new}) we are breaking a flavor symmetry: the modified theory has
$d-2$ non anomalous global charges: $U(1)_F \times U(1)_B^{d-3}$. 
The supergravity dual, if it exists, will have therefore an internal
manifold $H$ with isometry $U(1)^2$ and therefore the corresponding CY
cone cannot be toric. We will refer to these theories as $T^2$ theories,
since $T^2$ is the maximal torus of the isometry group. 

It is easy to generalize to the case when $H$, if the supergravity
dual exists, may have only isometry $U(1)$: in the gauge theory we have to
break both the original flavor symmetries. We add therefore to the
superpotential all mesons mapped to integer points of $\mathcal{C}^*$ lying on a
plane passing through $\vec{m}_0$, that is points of the form:
$\vec{m}_0 + k \,\vec{d}_1 + h \, \vec{d}_2$ with $k$, $h$ integers and
$\vec{d}_1$, $\vec{d}_2$ suitable independent vectors. Both for global
and R-symmetries we have to add the constraints:
\begin{equation}
\sum_{i=1}^d \left( \vec{d}_1 \cdot \vec{V}_i \right) a_i=0 \qquad
\sum_{i=1}^d \left( \vec{d}_2 \cdot \vec{V}_i \right) a_i=0
\label{newt1}
\end{equation}  
that preserve only the original $d-3$ baryonic symmetries. 
We will refer to these theories as $T^1$ theories.

We will assign generic coefficients to all mesons
appearing in the superpotential and study in concrete examples the
moduli space of the resulting quiver gauge theory. For instance it is
always possible to write down the moduli space of the quiver gauge
theory as an intersection of surfaces in a complex space $\mathbb
C^n$, where the complex variables correspond to mesons and the
equations are the relations in the chiral ring of mesons (in the
abelianized theory). Note that the moduli spaces in the modified
theories are always cones since at least one $\mathbb C^*$ action of the
original $\mathbb C^{*3}$ toric action survives on mesons.  
For some values of the vector(s) $\vec{d}$ (or $\vec{d}_1$,
$\vec{d}_2$) there exist suitable choices of the coefficients in the 
superpotential for which the moduli space of the gauge theory is 
three dimensional. As already explained we expect in these cases that
the supergravity duals of these theories are in the general class $AdS_5
\times H$, with $H$ the Sasaki-Einstein base of the cones of the
moduli space. 

Interestingly some information about the geometry of the new non
toric cones obtained with such constructions can be deduced by the
original theory. Consider for instance the $T^2$ case; 
the geometry has now a $\mathbb C^{*2}$ action. 
Holomophic functions over the complex cone have two charges under the $\mathbb
C^{*2}$ actions, and therefore they are mapped to integer points in
the plane; we will call $\mathcal{C}^*_{T^2}$ the minimal cone in the plane
whose integer points are possible charges of holomorphic
functions\footnote{It would be interesting to know whether $\mathcal{C}^*_{T^2}$
  is also the image of the momentum map of the K\"ahler cone.}. The
difference with the toric case $\mathcal{C}^*\equiv \mathcal{C}^*_{T^3}$ is that
$\mathcal{C}^*_{T^2}$ is two dimensional and integer points of $\mathcal{C}^*_{T^2}$ may
have multiplicities greater than one; the multiplicity is the number
of holomorphic functions that are mapped to that point. We can count
the holomorphic functions by looking at 
mesons in the quiver gauge theory. Since the quiver is the same,
mesons in the non toric theory are the same as in the original toric
theory; the difference is that now mesons mapped to points
$\vec{m}+k\,\vec{d}$ in the $\mathcal{C}^*$ cone of the toric theory have all
the same trial charge because of equations (\ref{psimap}),
(\ref{new}). 
Therefore $\mathcal{C}^*_{T^2}$ is simply the quotient of the cone
$\mathcal{C}^*$ of the toric theory with respect to direction
$\vec{d}$. We will write:
\begin{equation}
\mathcal{C}^*_{T^2}=\pi(\mathcal{C}^*)
\label{pi}
\end{equation}
where $\pi$ stands for the projection along $\vec{d}$: 
$\vec{m}\sim \vec{m}+k\vec{d}$.
The non trivial fact is that we can obtain also the
multiplicities of holomorphic functions (linearly independent mesons)
with fixed charges under the $\mathbb C^{*2}$ action by counting the number of
integer points of the polyhedral cone $\mathcal{C}^*$ of the toric theory
belonging to the same line with direction $\vec{d}$. We have checked
this in some concrete examples; indeed the introduction of new
superpotential terms modifies the F-term linear relations between
mesons. It is reasonable that the number of linearly independent
mesons is not modified, but we do not have a general proof of this
fact. We conjecture that the multiplicities of holomorphic functions
with assigned charges is always obtained in this way by the quotient
of the original $\mathcal{C}^*$ along direction $\vec{d}$ when the moduli space
is three dimensional\footnote{When the dimension of the moduli space
  is less than three we found in concrete examples that some chiral
  fields, and therefore some mesons, must be set to zero in order to
  satisfy the F-term equations; the counting of holomorphic functions
  in these cases is not so straightforward.}.

Note that if we want to have a finite number of holomorphic functions
with fixed charges the vector $\vec{d}$ must lie in the complementary
of the union of $\mathcal{C}^*$ and $-\mathcal{C}^*$, with $\mathcal{C}^*$ the image of the momentum
map for the toric case:
\begin{equation}
\vec{d} \in \mathbb R^3 \, \backslash \, (\mathcal{C}^* \cup (-\mathcal{C}^*))
\label{finite}
\end{equation} 
This condition is equivalent to the fact that lines $\vec{m}+k\,
\vec{d}$ contain a finite number of integer points of $\mathcal{C}^*$ 
(if $\vec{d}$ were in $\mathcal{C}^* \cup (-\mathcal{C}^*)$, say in $\mathcal{C}^*$, then for 
any $\vec{m}\in \mathcal{C}^*$ the whole half line $\vec{m}+k\vec{d}$, 
$k \geq 0$, would be in $\mathcal{C}^*$).

Another condition on the vector $\vec{d}$ follows from the fact that
the line through $\vec{m}_0=(0,0,1)$: $\vec{m}_0+k\,\vec{d}$ must pass
through at least another integer point in $\mathcal{C}^*$, since this line
represents superpotential terms we are adding to modify the toric
case. That is at least one of the points
$\vec{m}_1=\vec{m}_0+\vec{d}$, $\vec{m}_{-1}=\vec{m}_0-\vec{d}$
must lie in $\mathcal{C}^*$, say $\vec{m}_1$ (of course we can exchange $\vec{d}$ with
$-\vec{d}$). It follows that $\vec{m}_1$ must lie on a facet (or) edge
of $\mathcal{C}^*$: if it were in the interior of $\mathcal{C}^*$ (strictly positive
integer scalar products with all $\vec{V}_i$) then
$\vec{m}_1-\vec{m}_0=\vec{d}$ would be again in $\mathcal{C}^*$ ($\vec{m}_0$ has scalar
product $1$ with all vectors $\vec{V}_i$), but this is in
contradiction with (\ref{finite}). Therefore we can add to the
superpotential only mesons $\vec{m}_1$ and/or $\vec{m}_{-1}$ along facets of
$\mathcal{C}^*$ (mesons mapped to $\vec{m}+k\,\vec{d}$ with $|k|>1$ lie outside
$\mathcal{C}^*$).   

The case $T^1$ with a single $\mathbb C^*$ action is completely analogous: the 
cone of holomorphic functions $\mathcal{C}^*_{T^1}$ (and the image of the momentum map)
is a half-line and is obtained by making the
quotient of the polyhedral cone $\mathcal{C}^*$ for the toric case with respect
to the plane generated by the directions $\vec{d}_1$, $\vec{d}_2$:
\begin{equation}
\mathcal{C}^*_{T^1}=\Pi(\mathcal{C}^*)
\label{Pi}
\end{equation}
where $\Pi$ stands for the quotient along the plane generated by $\vec{d}_1$, $\vec{d}_2$:
$\vec{m} \sim \vec{m}+k\vec{d}_1+h\vec{d}_2$.
The number of integer points of $\mathcal{C}^*$ in these planes counts the number of
holomorphic functions in the non toric cone with assigned charge, when
the moduli space of the quiver gauge theory is three dimensional. 
Again mesons added to the superpotential of the toric theory must lie
on facets of $\mathcal{C}^*$, and in order to obtain finite multiplicities
condition (\ref{finite}) must hold for $\vec{d}_1$, $\vec{d}_2$, and
for all integer vectors in their plane. Consider the rational line $R$
passing through the origin and perpendicular to the plane of
$\vec{d_1}$, $\vec{d}_2$, and call $\vec{n}$ the primitive integer
vector generating $R$. Then the condition for having finite
multiplicities can be more easily restated as: 
\begin{equation}
\vec{n} \, \in \,\, \textrm{interior of:} \,\, \mathcal{C} \cup (-\mathcal{C})
\label{finitet1}
\end{equation}
obviously the case $\vec{n}\rightarrow -\vec{n}$ is the same, and
hence we can suppose $\vec{n}\cdot \vec{g}_i >0$, where $\vec{g}_i$
are the generators over integer numbers of $\mathcal{C}^*$. Integer points
$\vec{m}$ in $\mathcal{C}^*$ on the same plane perpendicular to $\vec{n}$ have
constant scalar product: $\vec{m} \cdot \vec{n}=k$, for some integer
$k$. To prove (\ref{finitet1}) suppose that there exist two generators
of $\mathcal{C}^*$, say $g_1$, $g_2$, such that: $\vec{g}_1 \cdot \vec{n}\geq 0$
and $\vec{g}_2 \cdot \vec{n}\leq 0$. Then every plane $\vec{m} \cdot
\vec{n}=k$ would contain also the points in $\mathcal{C}^*$: $\vec{m}+h \left[
  \left(\vec{g}_1 \cdot \vec{n} \right) \vec{g}_2- \left(\vec{g}_2
  \cdot \vec{n} \right) \vec{g}_1 \right]$ for any positive integer
$h$. Thus condition (\ref{finitet1}) is equivalent to having finite
multiplicities. 

With these rules for counting multiplicities, it is not difficult to
write down the generating functions (characters) for multiplicities of holomorphic functions also in 
the non toric cases. As discovered in \cite{MSY2} it is possible to deduce
from these characters the volume of the Sasaki manifold $H$ in function
of the Reeb vector. It is therefore possible to show that the volume
computed in this way matches the results of a-maximization according
to the predictions of AdS/CFT correspondence. We will give a proof of
this fact for the class of theories introduced here in Section \ref{proof}.
Now we will give some examples of the
construction we have just explained. 

\section{Examples: $(P)dP_4$ theories}
\label{pdp4ex}

We consider now some concrete examples of the construction
suggested in the previous Section starting from the known cases of
$(P)dP_4$ theories and extending them.

\begin{figure}
\begin{center}
\includegraphics[scale=0.6]{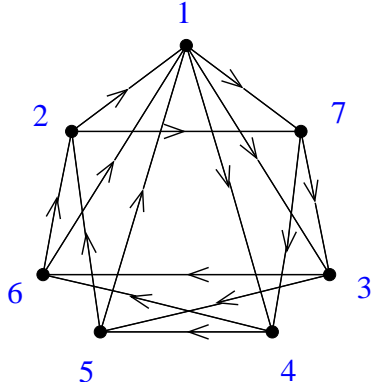} 
\caption{Quiver for $(P)dP_4$ theories.}
\label{quiverpdp4}
\end{center}
\end{figure}

The del Pezzo 4 surface, $dP_4$, is obtained by blowing up $\mathbb{P}^2$ at four
points at generic positions \footnote{that is no three of them are on
  a line. Therefore one can perform a $SL(3,\mathbb C)$ transformation
to put the four points in the fixed positions: $(1,0,0)$, $(0,1,0)$,
$(0,0,1)$, $(1,1,1)$ in $\mathbb{P}^2$. The moduli space of complex
deformations of $dP_4$ is a single point.}.
The complex cone over $dP_4$ can be endowed with a CY metric and the
dual quiver gauge theory is known in the literature; the quiver is
reported in Figure \ref{quiverpdp4}, the superpotential $W_{T^1}$ was found 
in \cite{wijnholt}. This CY has only one $\mathbb C^*$ action (the one 
corresponding to the complex fibration in the complex cone), that is
the maximal torus of isometry group is $T^1$. Correspondingly the dual
gauge theory has only one $U(1)_R$ and no non-anomalous flavor
symmetries. 

\begin{figure}
\begin{center}
\includegraphics[scale=0.6]{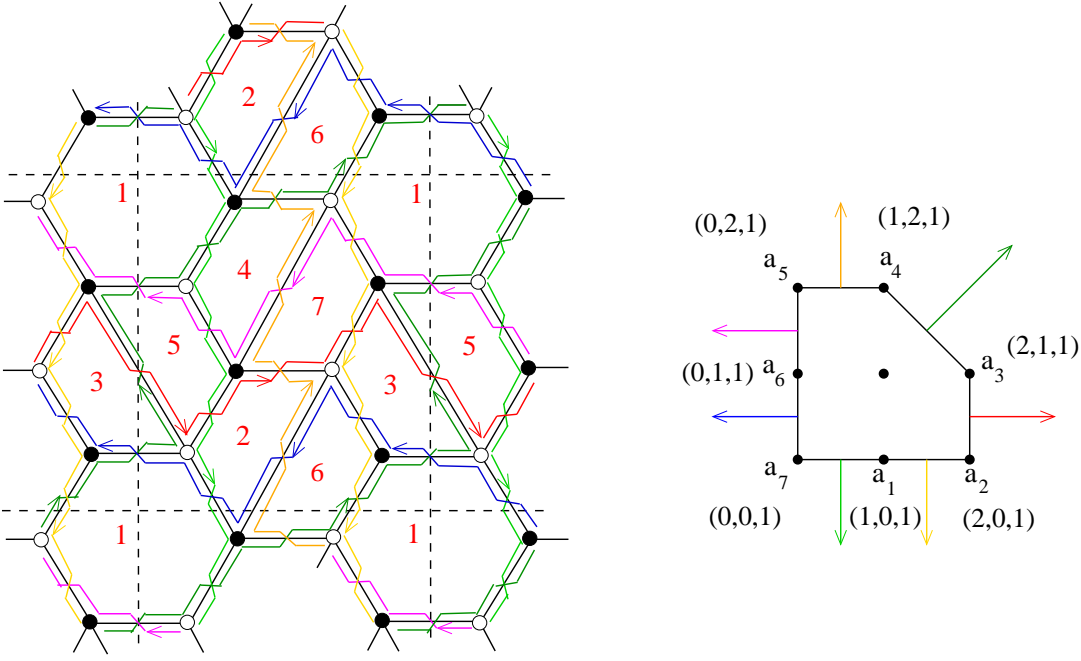} 
\caption{Dimer configuration and toric diagram for the toric $PdP_4$ theory.}
\label{pdp4}
\end{center}
\end{figure}

Instead if we choose as a K\"ahler basis of the CY a surface obtained
by blowing up $\mathbb{P}^2$ at non generic points it is possible to preserve
more symmetries. A possible example with isometry $T^3$ is the toric CY 
described by the toric diagram with vertices $V_i=(x_i,y_i,1)$: 
\begin{equation}
(0,0,1) \quad (1,0,1) \quad (2,0,1) \quad (2,1,1) \quad (1,2,1) \quad
  (0,2,1) \quad (0,1,1)
\label{directcone}
\end{equation}
We will refer to this manifold as the complex cone over Pseudo del
Pezzo 4, $PdP_4$, as already done in the literature. We draw the toric
diagram in Figure \ref{pdp4} where we report also the dimer of the
dual gauge theory. The dimer has $F=7$ gauge groups, $E=15$ chiral fields
and $V=8$ superpotential terms. 

Interestingly the quiver diagram of $PdP_4$ theory obtained from the
dimer is just the same as the quiver of $dP_4$ theory in Figure
\ref{quiverpdp4}: the difference between the two gauge theories is in
the superpotential. They have the same global baryonic symmetries (equal to
$d-3=4$ with $d$ the perimeter of the toric diagram), since they
depend only on the quiver diagram and not on the superpotential (the
total charge of every vertex and of every closed loop in the quiver is
zero for baryonic symmetries). But for $PdP_4$ the superpotential
allows two flavor symmetries, whereas for $dP_4$ new terms in the
superpotential are added to break both the original flavor
symmetries. As we will see we can interpret $dP_4$ theory as a
``quotient'' of $PdP_4$ in the sense of the previous Section.

\subsection{The toric case}
Let us start to study the toric theory.
In Figure \ref{pdp4} we draw zig-zag paths in the dimer corresponding
to vectors of the (p,q) web drawn in the same color; 
this correspondence allows to find the distribution of charges $a_i$ with
the method explained in \cite{proc}:
\begin{equation}
\begin{array}{l@{\rightarrow}l@{\quad}l@{\rightarrow}l@{\quad}l@{\rightarrow}l}
X_{17} & a_1+a_6+a_7 & X_{21} & a_7 & X_{27} & a_3+a_4\\
X_{73} & a_2 & X_{14} & a_1+a_2+a_3 & X_{74} & a_5\\
X_{13} & a_4+a_5+a_6 & X_{62} & a_5+a_6 & X_{51} & a_4+a_5\\
X_{61} & a_2+a_3 & X_{52} & a_1+a_2 & X_{36} & a_1+a_7\\
X_{45} & a_6+a_7 & X_{46} & a_4 & X_{35} & a_3
\end{array}
\label{rcar}
\end{equation}
where $X_{ij}$ is the chiral field from gauge group $i$ to $j$. The
parameters $a_i$ are associated with vertices of the toric diagram as
in Figure \ref{pdp4}. If the sum $a_1+a_2+\ldots a_7$ is equal to 2
(0) we get a parametrization of non anomalous R-symmetries (global
symmetries) in the toric theory.

We can start to study the closed loops in this
quiver, independently from the superpotential. There are 24
irreducible (that cannot be written as a product of smaller loops)
mesons:
\begin{equation}
\verb| | \hspace{-1.5em}%
\begin{array}{c}
\begin{array}{llll}
m_1= X_{61}X_{17}X_{74}X_{46} & m_3=X_{21}X_{13}X_{35}X_{52} &
m_5=X_{27}X_{73}X_{36}X_{62} & m_7=X_{14}X_{45}X_{51} \\
m_2=X_{51}X_{17}X_{73}X_{35} & m_4=X_{21}X_{14}X_{46}X_{62} &
m_6=X_{27}X_{74}X_{45}X_{52} & m_8=X_{13}X_{36}X_{61} \\ \\
n_1=X_{21}X_{13}X_{36}X_{62} & n_3=X_{17}X_{74}X_{45}X_{51} &
n_5=X_{27}X_{73}X_{35}X_{52} & n_7=X_{14}X_{46}X_{61} \\
n_2=X_{21}X_{14}X_{45}X_{52} & n_4=X_{17}X_{73}X_{36}X_{61} &
n_6=X_{27}X_{74}X_{46}X_{62} & n_8=X_{13}X_{35}X_{51} \\ \\
\end{array} \\ 
\begin{array}{l@{\qquad}l}
p_1=X_{21}X_{17}X_{74}X_{46}X_{62} &
q_1=X_{21}X_{17}X_{74}X_{45}X_{52} \\
p_2=X_{21}X_{17}X_{73}X_{35}X_{52} &
q_2=X_{21}X_{17}X_{73}X_{36}X_{62} \\
\end{array} \\[2em]
\begin{array}{l@{\qquad}l}
t_1=X_{27}X_{73}X_{35}X_{51}X_{14}X_{46}X_{62} &
t_2=X_{27}X_{73}X_{36}X_{61}X_{14}X_{45}X_{52}\\ 
t_3=X_{27}X_{74}X_{45}X_{51}X_{13}X_{36}X_{62} &
t_4=X_{27}X_{74}X_{46}X_{61}X_{13}X_{35}X_{52}
\end{array}
\end{array}
\label{mesdef}
\end{equation}
Since we are thinking to the abelianized theory with all $U(1)$ gauge
groups we will not write traces in front of mesons.

From these definitions and from (\ref{rcar}) we can deduce the
parametrization of charges $\Psi$ for mesons, and setting the charge
$\Psi$ equal to $\sum (\vec{m} \cdot \vec{V}_i)\, a_i$ as in
(\ref{psimap}) we can see to which point $\vec{m}$ of $\mathcal{C}^*$ each meson is mapped:
\begin{equation}
\begin{array}{lll@{\qquad}lll}
m_1, \ldots m_8 & \rightarrow & (0,0,1) & 
q_1, q_2 & \rightarrow & (-1,-1,3)\\
n_1, n_3, p_1   & \rightarrow & (-1,0,2) &
n_5, n_7 & \rightarrow & (1,0,0) \\
n_2, n_4, p_2 & \rightarrow & (0,-1,2) &
n_6, n_8 & \rightarrow & (0,1,0) \\[0.2em]
t_1, t_4 & \rightarrow & (1,1,0) &
t_2 & \rightarrow & (1,-1,2) \\
t_3 & \rightarrow & (-1,1,2) & & & 
\label{map}
\end{array}
\end{equation}
Mesons mapped to the same integer point in $\mathcal{C}^*$ are F-term equivalent
in the toric theory: the multiplicities of integer
points in $\mathcal{C}^*$ is equal to one in toric theories.
A basis for the cone $\mathcal{C}^*$, dual to the cone $\mathcal{C}$ in
(\ref{directcone}) is given by the points:
\begin{equation}
(0,1,0) \quad (-1,0,2) \quad (-1,-1,3) \quad (0,-1,2) \quad (1,0,0)
  \quad (0,0,1)  
\label{gen}
\end{equation} 
where the first five vectors are the perpendiculars to the facets of
$\mathcal{C}$ and we have to add also $(0,0,1)$ to have a basis over the
positive integer numbers.
Note that the inverse image of the six generators in (\ref{gen}) under
$\Psi$-map consists only of irreducible loops in the quiver, since
composite mesons are mapped to the sum of the points corresponding to
their constituents. Instead the irreducible mesons $t_1$, $t_2$, $t_3$
and $t_4$ are mapped to points in $\mathcal{C}^*$ that are not generators.

The superpotential for the toric theory can be read off from the
dimer and it can be written as:
\begin{equation}
W_{T^3}= \sum_{i=1}^8 c_i m_i
\label{wt3}
\end{equation}
Recall in fact that in the toric superpotential there appear only
mesons mapped to $\vec{m}_0=(0,0,1)$. We have also inserted general coefficients
in front of every meson $c_i$, $i=1, \ldots 8$. Many of them
can be reabsorbed with a rescaling of fields: $X_{ij} \rightarrow
r_{ij} X_{ij}$, under which also mesons are rescaled as $m_i
\rightarrow r_i m_i$ where the $r_i$ are products of the suitable
$r_{ij}$. In the space of mesons $m_1, \ldots m_8$ there is only one
relation following from the definitions (\ref{mesdef}):
\begin{equation}
m_1 m_3 m_5 m_7=m_2 m_4 m_6 m_8 \quad \textrm{or}: \quad
r_1 r_3 r_5 r_7=r_2 r_4 r_6 r_8
\label{rel}
\end{equation}
This equality implies that the ratio:
\begin{equation}
\frac{c_1 c_3 c_5 c_7}{c_2 c_4 c_6 c_8}
\label{ratio}
\end{equation}
is constant under field rescaling. Therefore since we have 8
coefficients in the superpotential (\ref{wt3}) and one relation
(\ref{rel}) we can reabsorb only 7 parameters in the superpotential
through rescaling. For instance we can put the superpotential in the
form:
\begin{equation}
W_{T^3}= m_1 + m_3 + m_5 + m_7
-b \left(m_2 + m_4 + m_6 + m_8 \right)
\label{wt32}
\end{equation}
where we are assuming that all the original coefficients $c_i$ are
different from zero. 

Now we have also to impose the F-term constraints: by solving the
F-term equations of (\ref{wt32}) it is easy to see that there exists a
complex three dimensional moduli space of vacua only if the ratio in
(\ref{ratio}) is equal to one, that is $b^4=1$. Of course any quartic
root of unity is equivalent, up to rescaling. With the choice $b=1$ we
find the usual result:
\begin{equation}
W_{T^3}= m_1 + m_3 + m_5 + m_7
- \left(m_2 + m_4 + m_6 + m_8 \right)
\label{wt33}
\end{equation}
In fact $m_1$, $m_3$, $m_5$, $m_7$ correspond to white vertices in the
dimer, whereas $m_2$, $m_4$, $m_6$, $m_8$ correspond to black vertices.
The analysis performed here is general for toric theories: all but one
coefficients in the superpotential can be reabsorbed through
rescaling, and F-term conditions imply that also the last parameter is
fixed as in (\ref{wt33}) if we want a three dimensional moduli space.
In fact we expect that for all toric three dimensional cones there are
no complex structure deformations that leave the manifold a complex cone. 
If $b^4 \neq 1$ in (\ref{wt32}) indeed the dimension of the moduli
space is reduced (there are one complex dimensional lines): 
this is a beta deformation of the toric theory \cite{conformal}.

\subsection{$T^1$ and $T^2$ examples}
\label{t1t2ex}
Now we are ready to modify the toric theory as explained in Section
\ref{mod}. We observe in fact that to obtain the gauge theory for
$dP_4$ we have to add to the superpotential mesons mapped to points
$\vec{m}_0 + k \, \vec{d}_1+ h \, \vec{d}_2$ of the toric $\mathcal{C}^*$ with:
\begin{equation}
\vec{d}_1=(1,0,-1) \qquad \vec{d}_2=(0,1,-1)
\label{d1d2}
\end{equation} 
The vectors $\vec{m}_0 + k \, \vec{d}_1+ h \, \vec{d}_2$ with $k$ and $h$
integers and belonging to $\mathcal{C}^*$ are exactly the generators of $\mathcal{C}^*$ in
(\ref{gen}), therefore the superpotential becomes:
\begin{equation}
W_{T^1}= \sum_{i=1}^8 a_i m_i+ \sum_{i=1}^8 c_i n_i+ 
\sum_{i=1}^2 \left( f_i p_i+g_i q_i \right)
\label{wt1}
\end{equation}
that is we can consider all the first 20 mesons in
(\ref{mesdef}), multiplied with generic coefficients $a_i$, $c_i$,
$f_i$, $g_i$; for simplicity we will consider the coefficients $a_i$ of
the toric terms different from zero. A particular choice of these
coefficients reproduces the theory for $dP_4$ \cite{wijnholt}.
In fact we can study the superpotential (\ref{wt1}) performing the
same analysis just explained in the toric case: solving F-term
equations and imposing the existence of a three-dimensional moduli
space we get some equations for the coefficients $a_i$, $c_i$, $f_i$,
$g_i$.
We have not studied in detail all the possible
branches of relations between these coefficients, but we
have explicitly verified that there are solutions admitting a three
dimensional cone of moduli space where the D3-branes can move.
The isometry for the non toric cone is $T^1$ (it is enough that a
suitable number of coefficients in (\ref{wt1}) is different from zero,
so that both original flavor symmetries are broken).
Once we have imposed that the moduli space is a three dimensional complex
cone, we have to remind that not all different choices of coefficients in the
superpotential (\ref{wt1}) 
determine inequivalent cones from the complex structure point of view. The
computation of the complex deformations of the non toric cone is non trivial.
In fact, differently from the toric
case, non toric cones may admit complex structure deformations that
leave the manifold a cone. One way to compute the deformations is to write the complex cone as
an intersection in some $\mathbb{C}^k$ space, where typically the complex
variables are associated with mesons, and then consider generic
linear redefinitions of the complex variables\footnote{In the toric case we
  saw that it is enough to consider rescalings of chiral fields and impose
  conditions for the existence of a three dimensional moduli space to
  reabsorbe all the coefficients in the superpotential, showing that there does
  not exists any complex structure deformation that leaves the manifold a
  cone. In more general cases one should consider all possible linear
  relations among mesons to get the correct counting of complex structure
  deformations. See Appendix B for further details.}.  
We have not performed explicitly the computation in the $T^1$ case starting
with generic coefficients in the superpotential (\ref{wt1}), however in this
case by simple geometrical considerations\footnote{In fact the $T^1$
  superpotential in \cite{wijnholt} for $dP_4$ belongs to this general class of
  superpotentials and it is known that the complex cone over $dP_4$ has no
  complex structure deformations that leave it a cone. Correspondingly the
  positions of the four blow-up points in $\mathbb{P}^2$ can be fixed with
  $SL(3,\mathbb{C})$ transformations, for instance to $(1,0,0)$, $(0,1,0)$, $(0,0,1)$
  and $(1,1,1)$.}  
we expect that there does not exist any complex structure deformation for
the $T^1$ cone that leaves it a cone.
 
With the same quiver, we can build gauge theories whose moduli space
is a non toric cone with isometry $T^2$: we can choose to project the
toric theory only along one direction in (\ref{d1d2}), for instance
choose\footnote{The other choice $\vec{d}=\vec{d}_2$ is equivalent up
  to a relabeling of gauge groups $3 \leftrightarrow 4$ and $5
  \leftrightarrow 6$ in the quiver.}: 
\begin{equation}
\vec{d}=\vec{d}_1=(1,0,-1)
\label{quotd1}
\end{equation} 
The points in $\mathcal{C}^*$ of the form $\vec{m}_0+k \, \vec{d}$ are:
$(-1,0,2)$, $(0,0,1)$, $(1,0,0)$ for the integers $k=-1,0,1$
respectively; the corresponding superpotential is:
\begin{equation} 
W_{T^2}= \sum_{i=1}^8 a_i m_i+ c_1 n_1 + c_3 n_3 + c_5 n_5 + c_7 n_7 +
f_1 p_1 
\label{wt2}
\end{equation}
Again it is simple to verify that there are choices of the
coefficients $a_i$, $c_i$, $f_1$ for which there exists a three dimensional
moduli space for the gauge theory with isometry $T^2$. Moreover in this $T^2$
case we explicitly checked that there are choices of parameters that give a
one complex dimensional family of complex
structure deformations that leave the manifold a cone\footnote{Interestingly
  this is in agreement with the expectation that these theories should describe complex
  cones over blow up of $dP_4$ at four points. In fact if the blow up points
  are not at generic positions, but they are chosen so as to preserve a $T^2$
  symmetry, then there may remain one complex parameter of
  deformations. Consider for instance the $T^2$ configuration of points on the
  same line: $(1,0,0)$, $(0,1,0)$, $(1,1,0)$, $(\alpha, \beta,0)$. Then the fourth point
  cannot be moved with $SL(3, \mathbb{Z})$ transformations 
  that keep fixed the first three points.}. 
As we will see all the cones in this family of complex
structure deformations have the same volume (for a
Sasaki-Einstein metric on their basis) and the same multiplicities for
holomorphic functions. In fact these features can be reconstructed
from the original toric theory and from the information about the
quotient (\ref{quotd1}).
For more details about the
moduli spaces of $(P)dP_4$ theories see Appendix B. 

\subsection{The cone of holomorphic functions}
\begin{figure}
\begin{center}
\includegraphics[scale=0.8]{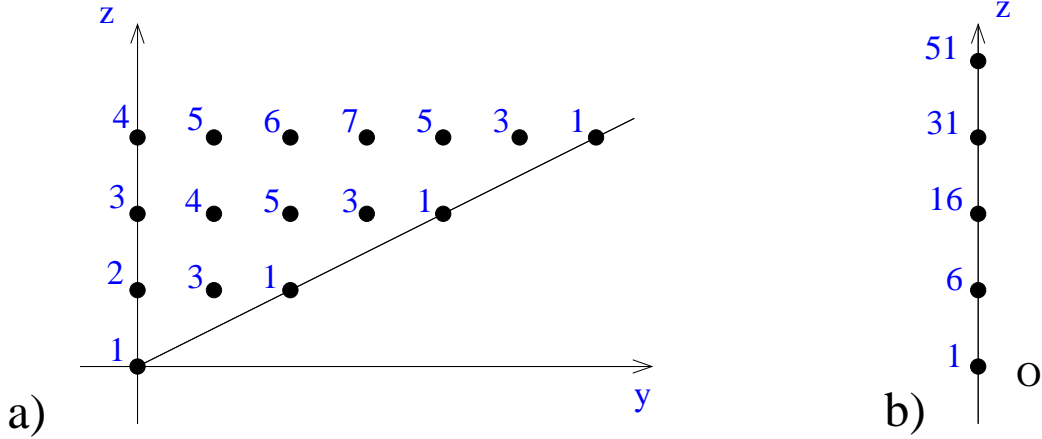} 
\caption{The multiplicities of holomorphic functions with fixed
  charge(s) over $(P)dP_4$ cones with a) isometry $T^2$, b) isometry
  $T^1$.}
\label{charge}
\end{center}
\end{figure}

As explained in Section \ref{mod}, it is easy to compute the possible
charges of holomorphic functions and the multiplicities (that is the
dimension of the vector space of holomorphic functions with an
assigned charge), by taking the quotient of the toric $\mathcal{C}^*$ along the
directions $\vec{d}_1,\vec{d_2}$ for $T^1$ theory or along $\vec{d}$
for $T^2$. Let us first of all find an $SL(3,\mathbb Z)$
transformation $A$ that sends the vector  $\vec{d}_1$, $\vec{d_2}$ into
$(1,0,0)$ $(0,1,0)$ respectively; choose for example:
\begin{equation}
A=\left(
\begin{array}{lll}
1 & 0 & 0 \\
1 & 2 & 1 \\
1 & 1 & 1
\end{array}
\right) \qquad \quad
\vec{d}^{\,\,'}_1=A \vec{d}_1=\left(
\begin{array}{l}
1 \\ 0 \\ 0
\end{array}
 \right), \quad
\vec{d}^{\,\,'}_2=A \vec{d}_2=\left(
\begin{array}{l}
0 \\ 1 \\ 0
\end{array}
 \right)
\label{matrixA}
\end{equation}
In this new system of coordinates the generating vectors (\ref{gen}) of
the toric $\mathcal{C}^*$ are sent respectively into:
\begin{equation}
(0,2,1) \quad (-1,1,1) \quad (-1,0,1) \quad (0,0,1) \quad (1,1,1)
  \quad (0,1,1)  
\label{genprimed}
\end{equation}
and the $\Psi$-map (\ref{map}) for the first 20 mesons in (\ref{mesdef}) becomes: 
\begin{equation}
\begin{array}{lll@{\qquad}lll}
m_1, \ldots m_8 & \rightarrow & (0,1,1) & 
q_1, q_2 & \rightarrow & (-1,0,1)\\
n_1, n_3, p_1   & \rightarrow & (-1,1,1) &
n_5, n_7 & \rightarrow & (1,1,1) \\
n_2, n_4, p_2 & \rightarrow & (0,0,1) &
n_6, n_8 & \rightarrow & (0,2,1) 
\end{array}
\end{equation}
Now the quotient along the direction
$\vec{d}^{\,\,'}=\vec{d}^{\,\,'}_1$ is simply the projection of the
cone $\mathcal{C}^*$ on the plane $(y',z')$; we draw it in Figure
\ref{charge}a).
This is the
cone of charges for the theories with isometry $T^2$ (\ref{wt2});
the multiplicities of holomorphic functions with charge $(n',m')$
are the number of points of $\mathcal{C}^*$ projected to $(n',m')$. 
For example among the
generators (\ref{genprimed}), the two vectors (-1,0,1), (0,0,1) are
mapped into (0,1); the three vectors (-1,1,1), (1,1,1), (0,1,1) are
mapped into (1,1) and the vector (0,2,1) is mapped into (2,1).

In the same way one can obtain the cone (indeed a half-line) of
charges of holomorphic functions for the theories with isometry $T^1$
(\ref{wt1}) by projecting the toric $\mathcal{C}^*$ on the axis $z'$; 
we draw it in Figure \ref{charge}b).
All the generators (\ref{genprimed}) are mapped into the point 1,
which has therefore multiplicity equal to 6.

We can perform some checks that the cones of holomorphic functions are
those drawn in Figures \ref{charge} by writing down the linear
relations among mesons induced by F-term relations in the different 
theories and counting the number of linearly independent mesons 
with a fixed charge. We will do that for mesons mapped to the
generators of $\mathcal{C}^*$, one should check the multiplicities also for more
complicated mesons.
Linear relations among mesons are obtained by multiplying the F-term
relation $\partial W / \partial X_{ij}=0$ with paths going from node
$i$ to node $j$. For example $X_{ij} \partial W / \partial
X_{ij}=0$ can be rewritten in terms of mesons
appearing in the superpotential; other linear relations can be deduced
manipulating non linear equations between mesons using these linear
constraints. 

In the toric case (\ref{wt33}) the linear relations are simple
equalities; for the first 20 generators in (\ref{mesdef}) we get:
\begin{equation}
\begin{array}{l@{\qquad}l}
m_1=\ldots =m_8 & q_1=q_2 \\
n_1=n_3=p_1 & n_5=n_7 \\
n_2=n_4=p_2 & n_6=n_8
\label{lineart3}
\end{array}
\end{equation}
More generally in toric theories mesons mapped to the same point in
$\mathcal{C}^*$ are equal, so that the cone of charges of holomorphic functions is $\mathcal{C}^*$
and each integer point in $\mathcal{C}^*$ has multiplicity equal to one.

Consider instead the theory $T^2$ (\ref{wt2}) and rescale the
coefficients in the superpotential to the form\footnote{one could
  perform also other rescalings. A possible choice of relations among
  coefficients that assures the existence of a three dimensional
  moduli space is: $e=(b^2 f)/(a^2-d f)$, $g=(a b^4-a^5+2 a^3 d f-a d^2
  f^2)/(b^3 f)$, $c=(a^4-b^4-a^2 d f)/(b^2 f)$.}:
\begin{equation}
W= a(m_1+m_3+m_5+m_7)-b(m_2+m_4+m_6+m_8)+c n_1 -d n_3+ e n_5 -f n_7 
+g p_1  
\end{equation}
The 13 mesons $m_1 \ldots m_8$, $n_1$, $n_3$, $p_1$, $n_5$, $n_7$, 
that are all mapped to the point $(1,1)$, satisfy a set of 10
independent linear relations:
\begin{equation} 
\begin{array}{lll}
a m_1 - b m_2 - d n_3 + g p_1=0 & a m_3-b m_4+c n_1+g p_1=0 & 
a m_5-b m_6+e n_5=0 \\
-b m_2+a m_5+e n_5=0 &  -b m_4+a m_7-f n_7=0 & a m_3-b m_8+c n_1=0 \\
-b m_4+a m_5+c n_1+g p_1=0 & -b m_2+a m_7-d n_3=0 & -f m_1+a n_3-b
p_1=0 \\
-f m_4-b n_1+a p_1=0 & & 
\label{lineart21}
\end{array}
\end{equation}
so that we have verified that there are 3 independent holomorphic
functions with charge $(1,1)$.
The 5 mesons $n_2$, $n_4$, $p_2$, $q_1$, $q_2$, that are mapped to
the point $(0,1)$ satisfy 3 independent linear relations:
\begin{equation}
\begin{array}{lll}
a p_2 - b n_4 +c q_2=0 & a n_2 -b p_2 -d q_1 & e p_2 - b q_1 + a q_2=0
\label{lineart22}
\end{array}
\end{equation}
so that the point $(0,1)$ has multiplicity 2.
The mesons $n_6$, $n_8$, mapped to $(2,1)$, satisfy one linear
relation:
\begin{equation}
\left( a^2-d f \right)n_6-b^2 n_8=0
\label{lineart23}
\end{equation} 
and therefore the point $(2,1)$ has multiplicity 1.
Note that all relations written for the theory $T^2$ reduce to
equalities in (\ref{lineart3}) if we set $c=d=e=f=g=0$, and $a=b=1$,
that is when we recover the toric superpotential. 
Note that obviously both in the toric theory $T^3$ and in the $T^2$
theory there are linear relations only for mesons having the same
global charges. The introduction of
superpotential terms that break one flavor symmetry modifies the
linear relations of the toric case: now all mesons mapped to aligned
integer points $\vec{m}+ k \vec{d}$ of the toric
$\mathcal{C}^*$ may appear in the same linear relations since they have the same
charges in the $T^2$ theory (\ref{new}). 

The case of the theories with symmetry $T^1$ is completely analogous:
linear relations written before (\ref{lineart21}), (\ref{lineart22}),
(\ref{lineart23}) 
are extended to include all the 20 considered 
mesons, that are now mapped to the same point 1 in Figure
\ref{charge}b); this point has therefore multiplicity equal to 6.

\begin{figure}
\begin{center}
\includegraphics[scale=0.6]{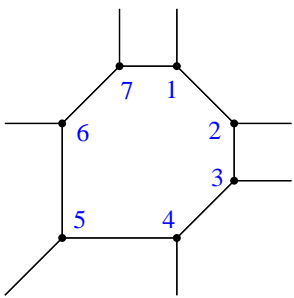} 
\caption{Resolution of complex cone over $PdP_4$}
\label{resolution}
\end{center}
\end{figure}

\subsection{Characters and volumes}
The equivariant index of the Cauchy-Riemann operator $\bar{\partial}$ \cite{AS}
on a (CY) cone allows to count the number $c_{m}$ of holomorphic functions
with fixed charge $m\equiv(m_1,\ldots,m_r)$ \cite{MSY2}, 
where $r$ is the dimension of
the maximal torus $T^r$ of the isometry group 
(in our examples $r=1,2$
or $3$). The character $C(q)$, $q\equiv(q_1,\ldots q_r)$ is the
generating function of the number of holomorphic functions:
\begin{equation}
C(q)=\sum c_m \, q^m \equiv \sum c_m \, q_1^{m_1} \ldots q_r^{m_r}
\end{equation}
For toric cones holomorphic functions are in one to one correspondence
with the set $\mathcal{S}_C$ of integer points of $\mathcal{C}^*$, and hence 
the character is:
\begin{equation}
C(q)=\sum_{m \in \mathcal{S}_C} q^m
\label{exp}
\end{equation}
moreover it can be easily computed by resolving
the cone \cite{MSY2}: the contributions to $C(q)$ come only from the
fixed points of the toric action.

In our example we can choose for $PdP_4$ the resolution drawn in
Figure \ref{resolution} with fixed points $p_A$, $A=1,\ldots 7$; at
each fixed point $A$ there are $n=3$ primitive edge vectors $u_i^{(A)}$,
$i=1,2,3$ (in the coordinates where the generators are (\ref{gen})):
\begin{equation}
\begin{array}{llll}
p_1: & u_1^{(1)}=(0,1,0) & u_2^{(1)}=(-1,0,1) & u_3^{(1)}=(1,-1,0)\\ 
p_2: & u_1^{(2)}=(1,0,0) & u_2^{(2)}=(0,-1,1) & u_3^{(2)}=(-1,1,0)\\
p_3: & u_1^{(3)}=(1,0,0) & u_2^{(3)}=(0,1,-1) & u_3^{(3)}=(-1,-1,2)\\
p_4: & u_1^{(4)}=(0,-1,2) & u_2^{(4)}=(-1,0,1) & u_3^{(4)}=(1,1,-2)\\
p_5: & u_1^{(5)}=(-1,-1,3) & u_2^{(5)}=(1,0,-1) & u_3^{(5)}=(0,1,-1)\\
p_6: & u_1^{(6)}=(-1,0,2) & u_2^{(6)}=(1,1,-2) & u_3^{(6)}=(0,-1,1)\\
p_7: & u_1^{(7)}=(0,1,0) & u_2^{(7)}=(-1,-1,2) & u_3^{(7)}=(1,0,-1)\\
\label{uvec}
\end{array}
\end{equation}
Then the character of the $PdP_4$ theory can be computed using the
general formula for toric cones \cite{MSY2}:
\begin{equation}
C_{T^3}(q)=\sum_{p_A} \prod_{i=1}^3 \frac{1}{1-q^{\, u_i^{(A)}}}
\label{ctoric}
\end{equation}
where again for the vectors $q$, $u$, the expression $q^u$ stands for:
$q_1^{u_1} q_2^{u_2} q_3^{u_3}$.

We perform the change of coordinates in (\ref{matrixA}) and apply equation
(\ref{ctoric}) after transforming vectors in (\ref{uvec}), that is
with the replacement $u_i^{(A)}\rightarrow A \, u_i^{(A)}$; after some
simplifications we get:
\begin{equation}
C_{T^3}(q)=\frac{q_1 \left(q_2 \left(-q_2^2 q_3^2-q_2
   q_3+q_2+1\right) q_3^2+q_1 \left(q_2^2 q_3^2+q_2
   (q_3-1) q_3-1\right)\right)}{(q_1-q_3) (q_3-1) \left(q_2^2
   q_3-1\right) \left(q_2 q_3 q_1^2-\left(q_2^2 q_3^2+1\right)
   q_1+q_2 q_3\right)}
\label{ct3}
\end{equation}

The cone of holomorphic functions for the theory $T^2$ is obtained by
projecting the toric cone $\mathcal{C}^*$ along the direction
$\vec{d}^{\,\,'}=(1,0,0)$; therefore the character $C_{T^2}(q)$ for
$T^2$ theories, $q=(q_2,q_3)$, is obtained simply by setting $q_1=1$ in
equation (\ref{ct3}), this is evident from the expansions in
(\ref{exp}). We obtain:
\begin{equation}
C_{T^2}(q)=\frac{-q_2^2 q_3^3-2 q_2 (q_3-1) q_3+1}{(q_3-1)^2 (q_2
   q_3-1) \left(q_2^2 q_3-1\right)}
\label{ct2}
\end{equation}
and this is the generating function for Figure \ref{charge}a).

Analogously the character $C_{T^1}(q)$ for the theory $T^1$, $q=q_3$, is obtained by
setting $q_1=1$ and $q_2=1$, since now the cone of charges for
holomorphic functions is obtained by projecting the toric $\mathcal{C}^*$ along
the plane generated by $\vec{d}^{\,\,'}_1=(1,0,0)$ and
$\vec{d}^{\,\,'}_2=(0,1,0)$. We obtain:
\begin{equation}
C_{T^1}(q)=\frac{q_3^2+3 q_3+1}{(1-q_3)^3}
\label{ct1}
\end{equation}
and this is the generating function for Figure \ref{charge}b). 

Remarkably in \cite{MSY2} it was shown that the volume of a
Sasaki metric over the base of the cone 
depends only on the Reeb vector $b=(b_1,\ldots
b_r)$, and can be expressed in terms of contributions localized on the
vanishing locus of the Reeb vector field; comparing their formula for
the volume with that of the equivariant index for the character, the
authors of \cite{MSY2} proved the interesting relation:
\begin{equation}
V(b)=\lim_{t \rightarrow 0} t^n C(e^{-t b})
\label{limit}
\end{equation}  
where in our case $n=3$ is the complex dimension of the cone over the
Sasaki manifold; the character is evaluated in $q=e^{-t b}$
defined as $q_i=e^{-t b_i}$. The function $V(b)$ is the normalized
volume function for the Sasaki manifold:
\begin{equation}
V(b) \equiv \frac{\textrm{Vol}(b)}{\textrm{Vol}(S^{2n-1})} 
\qquad \qquad \textrm{for} \,\,\, n=3: \,\,\,
V(b) =\frac{\textrm{Vol}(b)}{\pi^3} 
\end{equation} 

For toric manifolds the limit in (\ref{limit}) applied to
(\ref{ctoric}) yields:
\begin{equation}
V(b)=\sum_{p_A} \prod_{i=1}^3 \frac{1}{(b,u_i^{\,(A)})}
\end{equation} 

In our examples, performing the limit (\ref{limit}) for equations
(\ref{ct3}), (\ref{ct2}), (\ref{ct1}) we obtain:
\begin{eqnarray}
V_{T^3}(b) & = & \displaystyle \frac{2 b_2^2+7 b_3 b_2+5 b_3^2-b_1 (2 b_2+3
   b_3)}{(b_1-b_3) b_3 (2 b_2+b_3) \left(b_1^2-(b_2+b_3)^2\right)}
   \nonumber \\
V_{T^2}(b) & = & \displaystyle \frac{2 b_2+5 b_3}{b_3^2 (b_2+b_3) 
                 (2 b_2+b_3)} \nonumber \\ 
V_{T^1}(b) & = & \displaystyle \frac{5}{b_3^3}
\label{volumes}
\end{eqnarray}

These are the formulas for the normalized volume of a Sasaki metric of
Reeb vector $b$ over the basis of the cones we are considering.
Note that $V_{T^2}(b)$ is obtained from $V_{T^3}(b)$ by setting
$b_1=0$: in fact $q_1=e^{-t b_1}$ and $q_1=1$ to obtain the $T^2$
theory. In the same way $V_{T^1}(b)$ is obtained by setting $b_1=b_2=0$
in $V_{T^3}(b)$. 

We should try to find the position of the Reeb vector corresponding to
a Sasaki-Einstein metric to compute the volume in this case. Indeed
the Reeb vector of a Sasaki-Einstein metric is at the minimum of the
functions $V_{T^r}$ restricted to a suitable\footnote{this affine
  space is identified by the request: $\mathcal L_{r\partial /
    \partial r} \Omega=n\Omega$, with $\Omega$ a closed nowhere
  vanishing $(n,0)$ form.} affine space of dimension
$r-1$ \cite{MSY2}, and $b$ inside the direct cone (the dual of the
image of the momentum map). For toric cases \cite{MSY} this is the
plane $z=3$ in coordinates (\ref{directcone}); 
in our basis (\ref{volumes}) this plane becomes\footnote{Recall that
  if vectors in $\mathcal{C}^*$ transform as $m\rightarrow A m$, than vectors in
the direct cone $\mathcal{C}$, like the Reeb vector, transform as $n\rightarrow
{}^tA^{-1} n$, so that the scalar product $(n,m)$ is constant.}:
$b_2+b_3=3$. The Z-minimization for the toric case leads to: 
$\vec{b}=(-0.37908,-0.37908,3.37908)$.
In non toric theories the affine space where to perform
the volume minimization is not explicitly known; therefore in our
examples we will extract the position of the Reeb vector for
Sasaki-Einstein metrics from the gauge theory, studying the scaling
dimensions of mesons.

\subsection{Central and R-charges}
In all kinds of theories $T^1$, $T^2$, $T^3$ we are considering, the
trial R-charges for chiral fields can be parametrized with the $a_i$,
$i=1,\ldots 7$, associated with vertices of the toric diagram (Figure
\ref{pdp4}) as in (\ref{rcar}), but we have to impose more linear
constraints on them as the number of symmetries decreases. 
In fact condition (\ref{erre}) must always be imposed in all cases $T^1$, $T^2$, $T^3$:
\begin{equation}
a_1+a_2+a_3+a_4+a_5+a_6+a_7=2
\end{equation}
If the isometry is $T^2$ we have the further constraint (\ref{new}), which in our case is:
\begin{equation}
a_2 + a_3 - a_5 - a_6 - a_7 = 0
\end{equation}
and if the isometry is $T^1$ there are two linear constraints (\ref{newt1}), which in our example are:
\begin{equation}
\begin{array}{l}
a_2 + a_3 - a_5 - a_6 - a_7 = 0\\
-a_1 - a_2 + a_4 + a_5 - a_7 = 0
\end{array}
\end{equation}

The trial central charge $a$ can be written as:
\begin{equation}
\begin{array}{lll}
a & = & \frac{9}{32} \textrm{tr} \, R^3 = 
\frac{9}{32} \left[ 7 + \left( a_1 + a_6 + a_7 -1 \right)^3
+ \left( a_7 -1 \right)^3 +  \left( a_3 + a_4 -1 \right)^3 
+ \left( a_2 -1 \right)^3 \right. \\
 & & + \left( a_1 + a_2 + a_3 -1 \right)^3
+ \left( a_5 -1 \right)^3 + \left( a_4 + a_5 + a_6 -1 \right)^3
+ \left( a_5 + a_6 -1 \right)^3 \\
& & + \left( a_4 + a_5 -1 \right)^3
+ \left( a_2 + a_3 -1 \right)^3 + \left( a_1 + a_2 -1 \right)^3
+ \left( a_1 + a_7 -1 \right)^3 \\
& & + \left( a_6 + a_7 -1 \right)^3
+ \left( a_4 -1 \right)^3 + \left( a_3 -1 \right)^3
\left. \right]
\end{array}
\label{afun}
\end{equation}
where we have used (\ref{rcar}) for the chiral fields. Note moreover
that the trace of the trial R-symmetry is zero $\textrm{tr} \, R=0$ in
all (non) toric theories built modifying toric theories as in Section
\ref{mod}. The proof is the same as in toric theories \cite{aZequiv},
since the chiral field content and relation (\ref{erre}) are the same
also in the modified non toric theories.  

Now we can perform a-maximization \cite{intriligator} and find exact
R-charges; the maximization of (\ref{afun}) is performed on an affine
space of dimension 6 for our $T^3$ theory, of dimension 5 for the
$T^2$ theories and of dimension 4 for the $T^1$ theories. The results
are\footnote{In the toric case the exact R-charges are roots of cubic
  equations, we give only numerical values.}: 
\begin{equation}
\begin{array}{l}
\begin{array}{lllll}
\textrm{case} \,\, T^3: & a_2=a_5\simeq 0.42698 
& a_3=a_4\simeq 0.29777 & a_7\simeq 0.55049 & a_1=a_6=0\\[1.5em]
\textrm{case} \,\, T^2: & a_2=a_7=\displaystyle\frac{2}{3}(\sqrt{3}-1)
& a_3=a_5=\displaystyle\frac{4}{3}(2-\sqrt{3}) 
& a_4=\displaystyle\frac{4}{\sqrt{3}}-2 & a_1=a_6=0
\end{array}\\[2.5em]
\,\,\,\textrm{case} \,\, T^1: \,\,\,\, 
a_2=a_3=a_4=a_5=a_7=\displaystyle\frac{2}{5} \,\,\,\,\,\,\, a_1=a_6=0 
\end{array}
\end{equation} 
and the values of central charges $a$ for the different theories are:
\begin{equation}
a_{T^3}=1.41805 \qquad a_{T^2}=6\sqrt{3}-9 \qquad a_{T^1}=\frac{27}{20}
\label{avalues}
\end{equation}

Using the values above we can compute the exact R-charges $R$ of
mesons and hence the scaling dimensions $\Delta=3/2 R$, which can be
also expressed through the Reeb vector $\vec{b}$ of a Sasaki-Einstein
metrics: $\Delta=\vec{b}\cdot \vec{m}$. Considering the mesons mapped
to generators $\vec{m}$ of the cone of charges for holomorphic
functions, we are able to find the Reeb vector corresponding to a
Sasaki-Einstein metric. The results, in the same coordinate system of
(\ref{volumes}) are: 
\begin{equation}
\begin{array}{ll}
\textrm{case} \,\, T^3: & \vec{b}=(b_1,b_2,b_3)=(-0.37908,-0.37908,3.37908)\\
\textrm{case} \,\, T^2: & \vec{b}=(b_2,b_3)=(3-2\sqrt{3},2\sqrt{3})\\
\textrm{case} \,\, T^1: & \vec{b}=b_3=3\\
\end{array}
\label{reebti}
\end{equation}
Note that in the toric case the Reeb vector lies in the plane
$b_2+b_3=3$ as expected and the result here agrees with that of
Z-minimization, as already proved in the general toric case. The case
$T^2$ is less trivial since we have three generators of the cone of
charges: $\vec{m}=(0,1)$, $\vec{m}=(1,1)$ and $\vec{m}=(2,1)$. The
three equations $\Delta=\vec{m}\cdot \vec{b}$ in the case $T^2$ are
respectively: 
\begin{equation}
b_3=2\sqrt{3} \qquad \quad b_2+b_3=3 \qquad \quad 2 b_2+b_3=6-2\sqrt{3}
\end{equation}
This system in two unknowns is indeed consistent and the solution is
in (\ref{reebti}). 

Inserting the values for the Reeb vectors (\ref{reebti}) into the
expressions for the volumes (\ref{volumes}) we are able to find the
normalized volumes of the basis of the CY cones: 
\begin{equation}
V_{T^3}=0.176299 \qquad V_{T^2}=\frac{3+2\sqrt{3}}{36} \qquad V_{T^1}=\frac{5}{27}
\label{Vvalues}
\end{equation}
Note that our methods allow to compute the volumes also for non
complete intersections.
Now it is possible to compare these volumes with the values for the
central charge $a$ (\ref{avalues}); according to AdS/CFT predictions
the following relation must hold: 
\begin{equation}
a=\frac{\pi^3}{4 \textrm{Vol}}=\frac{1}{4 V}
\label{matching}
\end{equation}
and it is easy to check that (\ref{avalues}) match (\ref{Vvalues}) in
all three cases $T^3$, $T^2$ and $T^1$. A general proof of this
matching for toric theories was given in \cite{aZequiv}; in Section
\ref{proof} we will give a simple proof of (\ref{matching}) for the
class of non toric theories obtained by modifying toric theories as in
Section \ref{mod}. 

\section{Generalized conifolds of type $A_k$}
\label{lop}
In this Section we apply the ideas of Section \ref{mod} to another
non-toric theory already known in the literature \cite{Gubser,
  Esperanza}, that we will call generalized conifold of type $A_2$. 
This theory is part of a big family of $\mathcal{N}=1$ superconformal
quiver gauge theories that are infrared fixed points of the
renormalization group flow induced by deforming $ADE$ $\mathcal{N}=2$
superconformal field theories by mass terms for the adjoint chiral
fields \cite{Gubser, Esperanza}. We will extend our analysis also to
the generalized conifolds of type $A_k$. 

\begin{figure}[h!]
\centering
\includegraphics[scale=0.7]{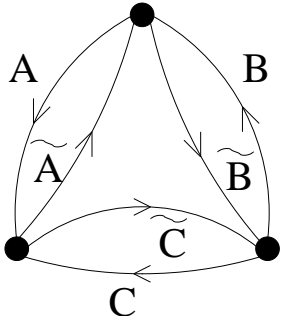}
\caption{The quiver of the generalized conifold of type $A_2$.}\label{lopez3}
\end{figure}

The generalized conifold of type $A_2$ is the $\mathcal{N}=1$
superconformal quiver gauge theory that lives on a stack of $N$
parallel $D3$-branes at the singular point of a type $D_4$ complex
three-fold \cite{Arnold}: 
\begin{equation}\label{D4}
x^3 + y^2 x = z w
\end{equation}
The theory has the gauge group $ SU(N) \times SU(N) \times SU(N) $ and
six chiral matter fields $A$, $\tilde A$, $B$, $\tilde B$, $C$,
$\tilde C$ that transform under the gauge group as shown in Figure
\ref{lopez3} \cite{Esperanza}. 
The superpotential is given by:
\begin{eqnarray}\label{l3supot}
W = a_1^2  \hbox{ Tr }(A\tilde A )^2 + a_2^2  \hbox{ Tr }(B\tilde B )^2  + a_3^2  \hbox{ Tr }(C\tilde C )^2 + \nonumber \\ 2 a_1 a_3\hbox{ Tr }A\tilde A \tilde C C +
2 a_1 a_2  \hbox{ Tr }B\tilde B \tilde A A  +  2 a_2 a_3  \hbox{ Tr }C\tilde C \tilde B B  
\end{eqnarray}
where $a_1$, $a_2$, $a_3$ are three independent coupling constants.
To reconstruct the geometry from the field theory we reduce as usual
to the abelian case in which the gauge group is $U(1)^3 $ and we
define the following gauge invariants:  
\begin{eqnarray}\label{x123wz}
x_1 = A \tilde A, \quad x_2 = B \tilde B, \quad  x_3 = C \tilde C \nonumber \\
z = A B C  \quad  w = \tilde A \tilde B \tilde C 
\end{eqnarray}
These variables are subject to the constraint:
\begin{equation}\label{con1}
x_1 x_2 x_3 = z w
\end{equation}
and in term of these gauge invariants the superpotential (\ref{l3supot}) can be written as
\begin{equation}\label{spotx}
W=(a_1 x_1 + a_2 x_2 + a_3 x_3)^2 
\end{equation} 
The $F$-term equations reduce to the linear constraint: $a_1 x_1+a_2
x_2+a_3 x_3=0$, hence we can use $x_1$, $x_2$, $z$, $w$ as the
generators of the mesonic chiral ring and equation (\ref{con1})
becomes, after fields rescaling, $x_1 x_2 (x_1+x_2)=z w$. This is
equivalent to the cone in (\ref{D4}) after the complex variables
redefinitions: $x_1 = (x + i y)/\sqrt[3]{2}$, $x_2= (x
-iy)/\sqrt[3]{2}$. In this way the field theory reconstructs the
geometry along its Higgs branch \cite{Esperanza}. 

This theory is important for our task because it is the gauge theory
dual to a non-toric geometry. 
If we give arbitrary weights under a $\mathbb{C}^*$ action to the four
embedding variables of the singularity it is easy to show that
equation (\ref{D4}) admits only two independent actions. This imply
that the variety is of type $T^2$. The field theory has indeed only
two non-anomalous $U(1)$ symmetries that are of non-baryonic type
\cite{Esperanza} and they correspond to the imaginary part of the
$\mathbb{C}^*$ action; these symmetries can be organized as the
$R$-symmetry $U(1)_R$ and the flavor symmetry $U(1)_F$:  
\begin{equation}
\begin{array}{l|llllll}
       & A & \tilde A & B & \tilde B & C & \tilde C \\ \hline
U(1)_R & 1/2 & 1/2 & 1/2 & 1/2 & 1/2 & 1/2  \\[0.5em]
U(1)_F & 1 & -1 & 0 & 0 & 0 & 0
\end{array}
\end{equation}

The quiver gauge theory we have just discussed is the smallest one of
the infinite family of generalized conifolds of type $A_{k-1}$
\cite{Gubser, Esperanza}. The quivers of these theories can be
obtained from the associated affine Dinkin diagrams of the $A_{k-1}$
orbifold singularities by deleting the arrows of the adjoint chiral
superfields: there are $k$ $SU(N)$ gauge groups labeled by a periodic
index $i=1,\ldots k$ and $2k$ chiral superfields, divided in the two
sets $X_i$, and $\tilde X_i$. The chiral field $X_i$ goes from node
$i$ to $i+1$; instead $\tilde X_i$ goes from node $i+1$ to node
$i$. The quiver in Figure \ref{lopez3} corresponds to the case $k=3$. 
In the abelian case the theory has $U(1)^{k}$ gauge group and we can
define the minimal set of gauge invariants: 
\begin{equation}\label{gaugeinvk}
x_i = X_i \tilde{X}_i , \quad z = X_1 X_2 ... X_{k}, \quad w = \tilde{X_1}\tilde{X_2}...\tilde{X}_{k}. 
\end{equation} 
These $k+2$ mesons satisfy the following relation:
\begin{equation}\label{conk}
x_1 x_2...x_{k} = z w  
\end{equation} 
The superpotential for these theories written in term of the above mesons is:
\begin{equation}\label{supotk}
W = \sum_{i=1}^{k} ( a_i x_i ^2 + 2 b_i x_i x_{i-1})  
\end{equation} 
The study of the moduli space for these theories has already been
performed in \cite{Esperanza}: the F-term relations following from
(\ref{supotk}) reduce to the $k$ linear relations: $b_i x_{i-1}+a_i
x_i+b_{i+1}x_{i+1}=0$. In order to have a three dimensional cone, we
have to impose suitable relations on the parameters $a_i$, $b_i$ (see
equation (4.8) in \cite{Esperanza}) in the superpotential such that
only $k-2$ of the F-term linear relations are independent: we can use
them to express $x_3$, \ldots $x_k$ as linear functions of $x_1$ and
$x_2$. Hence relation (\ref{conk}) becomes, using also fields
rescaling \cite{Esperanza}:   
\begin{equation}\label{Ak}
x_1 x_2 (x_1 + x_2) \prod _{j=1}^{k-3} ( x_1 + \alpha _j x_2 )= z w 
\end{equation}
This equation in term of independent mesons $x_1$, $x_2$, $z$, $w$
expresses the three-dimensional cone as a complex submanifold in
$\mathbb C^4$. 
The $\alpha _j$ in (\ref{Ak}) cannot be reabsorbed into linear redefinitions of
the variables and hence parametrize the complex structure deformations
that leave the moduli space a cone. 

Looking at the $\mathbb{C}^*$ action on the embedding variables one
can show that the varieties (\ref{Ak}) allow only a $(\mathbb{C}^*)^2$
action. The superpotential of the theories (\ref{supotk}) is a quartic
polynomial in the elementary fields and, using the symmetries of the
quiver gauge theory, we immediately find that the exact $R$-charges of
the chiral superfields are all equal to $1/2$. Hence the results of
a-maximization are simple: 
\begin{equation}
\textrm{a-max:} \qquad R(X_i)=R(\tilde X_i)=\frac{1}{2} \qquad 
a=\frac{9}{32}\left[ k+k(-\frac{1}{2})^3+k(-\frac{1}{2})^3 \right]
=\frac{27}{128}k
\label{amaxak}
\end{equation} 

\begin{figure}
\begin{center}
\includegraphics[scale=0.6]{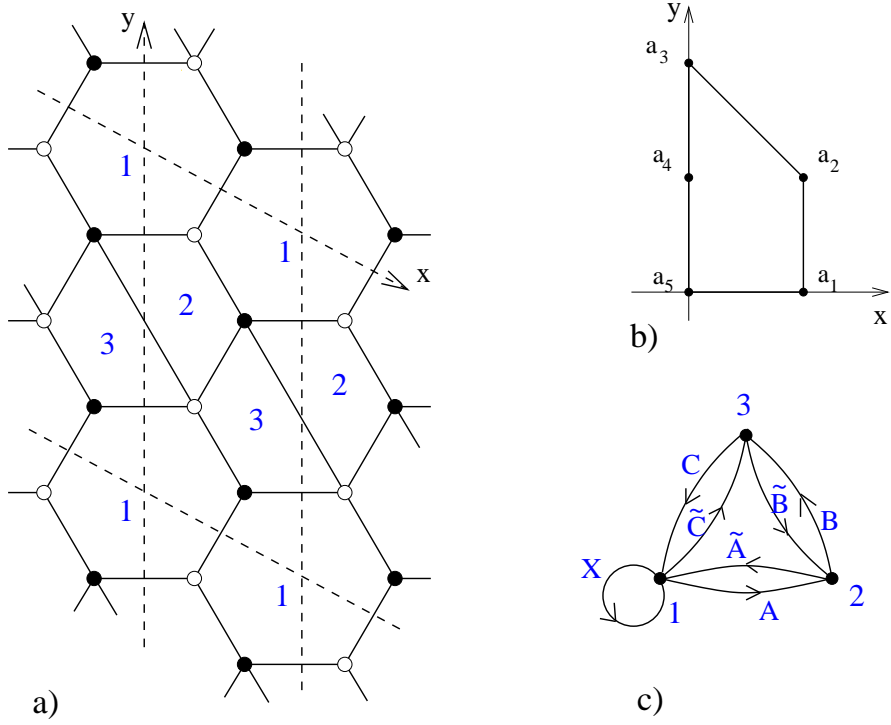} 
\caption{The SPP theory: a) dimer configuration b) toric diagram c) quiver diagram.}
\label{spp}
\end{center}
\end{figure}

Interestingly the generalized conifolds of type $A_{k-1}$ belong to
the class of non toric theories introduced in Section \ref{mod}. Let
us start with the case $k=3$; first of all we have to find a toric
theory with a quiver that can be reduced to that of Figure
\ref{lopez3}: we can choose for instance the well know theory of the
SPP, whose toric diagram $\mathcal{C}$ is: 
\begin{equation}
(0,0,1) \quad (1,0,1) \quad (1,1,1) \quad (0,2,1)
\end{equation}
The dimer, toric diagram and quiver are drawn in Figure \ref{spp}.
There are $F=3$ gauge groups, and $E=7$ chiral fields: there are all
the fields appearing in the generalized $A_2$ conifold: $A$, $\tilde
A$, $B$, $\tilde B$, $C$, $\tilde C$ plus an adjoint: $X_{11}\equiv
X$, to which we will have to give mass. In the SPP quiver the minimal
loops are the mesons in (\ref{x123wz}) plus the adjoint $X$.  
The toric superpotential can be written as:
\begin{equation}
W_{T^3}=x_1 X+x_2 x_3-x_3 X-x_1 x_2
\end{equation}

The dual cone $\mathcal{C}^*$ has four generators over integer numbers:
\begin{equation}
(0,1,0)  \quad (-1,0,1) \quad (-1,-1,2) \quad (1,0,0) 
\label{csspp}
\end{equation}
It is easy to find the $\Psi$-map for the generating mesons: $x_1$, $x_2$, $x_3$, $z$, $w$ and $X$:
\begin{equation}
\begin{array}{lll}
\textrm{meson} & \textrm{charge} & \Psi\textrm{-map}\\
w & a_2+2 a_3 +a_4 & (0,1,0)\\
x_1, x_3 & a_3+a_4+a_5 & (-1,0,1)\\
z & a_1+a_4+2 a_5 & (-1,-1,2)\\
x_2, X & a_1+a_2 & (1,0,0)
\end{array}
\label{psispp}
\end{equation}
where the $a_i$ are associated with vertices of the toric diagram as
in Figure \ref{spp}b). From the above table it is immediate to see
that we can choose as linearly independent mesons: $x_1$, $x_2$, $z$
and $w$ and that the SPP can be expressed in terms of this variables
as the surface in $\mathbb C^4$: $x_1^2 x_2=z w$, to be compared with
the equation for the generalized conifold of type $A_2$ (\ref{Ak}):
$x_1^2 x_2 + x_1 x_2^2=z w$. 
To modify the SPP superpotential and recover the generalized conifold
of type $A_2$ we have to introduce the mass term $X^2$, which is
mapped to the point of $\mathcal{C}^*$: $(2,0,0)$, as can be deduced
from (\ref{psispp}). 
We deduce that $\vec{d}=(2,0,0)-(0,0,1)=(2,0,-1)$, and that we can add
to the superpotential all mesons that are mapped to the three points
of $\mathcal{C}^*$: $(0,0,1)$, $(2,0,0)$, $(-2,0,2)$, that are all the
points of $\mathcal{C}^*$ of the form: $\vec{m}_0+k \vec{d}$. The
resulting modified superpotential is thus: 
\begin{equation}
W_{T^2}=\left( x_1 X+x_2 x_3+x_3 X+x_1 x_2 \right) 
+ \left( x_2^2 + X^2 +x_2 X \right)
+ \left( x_1^2 + x_3^2+ x_1 x_3 \right)
\end{equation} 
where in front of each term there is a generic coefficient that we
have understood for simplicity. Integrating out the massive field $X$,
one recovers the same superpotential of the generalized conifold of
type $A_2$ (\ref{spotx}). Therefore we have shown that this non toric
theory can be obtained by quotienting SPP along $\vec{d}=(2,0,-1)$.  

\begin{figure}
\begin{center}
\includegraphics[scale=0.6]{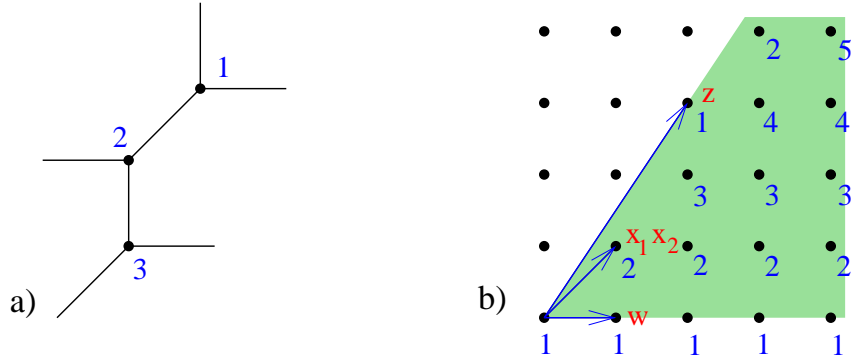} 
\caption{a) Resolution of the SPP singularity. b)
  $\mathcal{C}^*_{T^2}$ for the generalized conifold of type $A_2$
  with multiplicities of holomorphic functions.} 
\label{sppres}
\end{center}
\end{figure}

The character $C_{T^3}(q)$ for the SPP theory can be deduced easily
from the general formula (\ref{ctoric}) \cite{MSY2} and using the
following resolution of the SPP singularity: 
\begin{equation}
\begin{array}{llll}
p_1: & u_1^{(1)}=(0,1,0) & u_2^{(1)}=(1,0,0) & u_3^{(1)}=(-1,-1,1)\\ 
p_2: & u_1^{(2)}=(-1,0,1) & u_2^{(2)}=(0,-1,1) & u_3^{(2)}=(1,1,-1)\\
p_3: & u_1^{(3)}=(1,0,0) & u_2^{(3)}=(0,1,-1) & u_3^{(3)}=(-1,-1,2)
\end{array}
\label{uispp}
\end{equation}
which is shown in Figure \ref{sppres}a). But before computing the
character we perform an $SL(3,\mathbb Z)$ transformation $A$ that
sends $\vec{d}\rightarrow (0,0,1)$; we can choose for instance: 
\begin{equation}
A=\left(
\begin{array}{lll}
1 & 1 & 2\\
1 & 0 & 2\\
0 & 0 & -1
\end{array}
\label{aspp}
\right)
\end{equation}
with this change of coordinates the vectors of $\mathcal{C}^*$ (\ref{csspp}) become respectively:
\begin{equation}
(1,0,0) \quad (1,1,-1) \quad (2,3,-2) \quad (1,1,0)
\label{csnew}
\end{equation}
The cone $\mathcal{C}^*_{T^2}$ in this new system of coordinates is
the projection of $\mathcal{C}^*$ (\ref{csnew}) on the plane $(x,y)$,
since now $\vec{d}=(0,0,1)$; we see therefore that $w$ is mapped to
the point $(1,0)$, $z$ is mapped to $(2,3)$ and $x_1$, $x_2$, $x_3$
are mapped to $(1,1)$. The cone $\mathcal{C}^*_{T^2}$ for the
generalized conifold of type $A_2$ is drawn in Figure \ref{sppres}b). 

After applying the matrix $A$ (\ref{aspp}) to the vectors $u_i$ in
(\ref{uispp}) we find the character for the SPP in the new system of
coordinates \footnote{The character for complete intersections like
  SPP or the $D_4$ singularity can be 
also computed with simple methods directly from the equation. We thank
A. Hanany for an interesting discussion on this point}: 
\begin{equation}
C_{T^3}(q)=\frac{{q_3}^3-\left( {q_1}^3\,{q_2}^3\,q_3 \right)}
  {\left( 1 - q_1 \right) \,\left( 1 - q_1\,q_2 \right) \,\left( q_1\,q_2 - q_3 \right) \,
    \left( {q_1}^2\,{q_2}^3 - {q_3}^2 \right) }
\end{equation}
and the character for the generalized conifold of type $A_2$ is
obtained by inserting $q_3=1$ in $C_{T^3}$, in order to count all the
points of the toric $\mathcal{C}^*$ that are mapped to the same point
in the plane $(x,y)$: 
\begin{equation}
C_{T^2}(q)=\frac{1 - {q_1}^3\,{q_2}^3}
  {\left( 1 - q_1 \right) \,{\left( 1 - q_1\,q_2 \right) }^2\,\left( 1 - {q_1}^2\,{q_2}^3 \right) }
\end{equation}
The function $C_{T^2}(q)$ generates the multiplicities that are
reported in Figure \ref{sppres}b) and that can be computed directly in
some simple cases in the gauge theory of the generalized conifold of
$A_2$ type, in order to check our hypotheses for counting
multiplicities. The multiplicities of generators have already been
checked ($x_3$ is a linear combination of $x_1$ and $x_2$). Consider
for example the point $(3,3)$ with multiplicity $4$ according to
Figure \ref{sppres}b): with $x_1$, $x_2$, $z$ and $w$ there are five
mesons that can be mapped to $(3,3)$, they are $x_1^3$, $x_2^3$,
$x_1^2 x_2$, $x_1 x_2^2$, $z w$, but as we see from (\ref{con1}) $z w$
is a linear combination of $x_1^2 x_2$ and $x_1 x_2^2$, so that the
multiplicity is $4$. 

By performing the limit (\ref{limit}) for $C_{T^2}(q)$, or
equivalently for $C_{T^3}(q)$ and then setting $b_3=0$, we get the
following formula for the normalized volume of the 5d basis endowed
with a Sasaki metric with Reeb vector $(b_1,b_2)$: 
\begin{equation}
V_{T^2}(b_1,b_2)=\frac{3}{b_1\,\left( b_1 + b_2 \right) \,\left( 2\,b_1 + 3\,b_2 \right) }
\label{volt2spp}
\end{equation}

The position of the Reeb vector for a Sasaki-Einstein metric can be
computed again using the gauge theory and the formula
$\Delta=\vec{m}\cdot \vec{b}$: for the generating mesons mapped to the
points $(1,0)$, $(1,1)$, and $(2,3)$ in Figure \ref{sppres}b) we get
the three equations: 
\begin{equation}
b_1=\frac{9}{4} \quad b_1+b_2=\frac{3}{2} \quad 2 b_1+3 b_2=\frac{9}{4}
\end{equation}
that have the consistent solution: $(b_1,b_2)=(9/4,-3/4)$; this is a
non trivial check that scaling dimensions of mesons depend only on
their charge and on the Reeb vector. 
Inserting this value for the Reeb vector into (\ref{volt2spp}) we get
for the normalized volume $V=32/81$, that matches 
the value $a=81/128$ from a-maximization (\ref{amaxak}) according to
the predictions of AdS/CFT (\ref{matching}). 

\begin{figure}
\begin{center}
\includegraphics[scale=0.5]{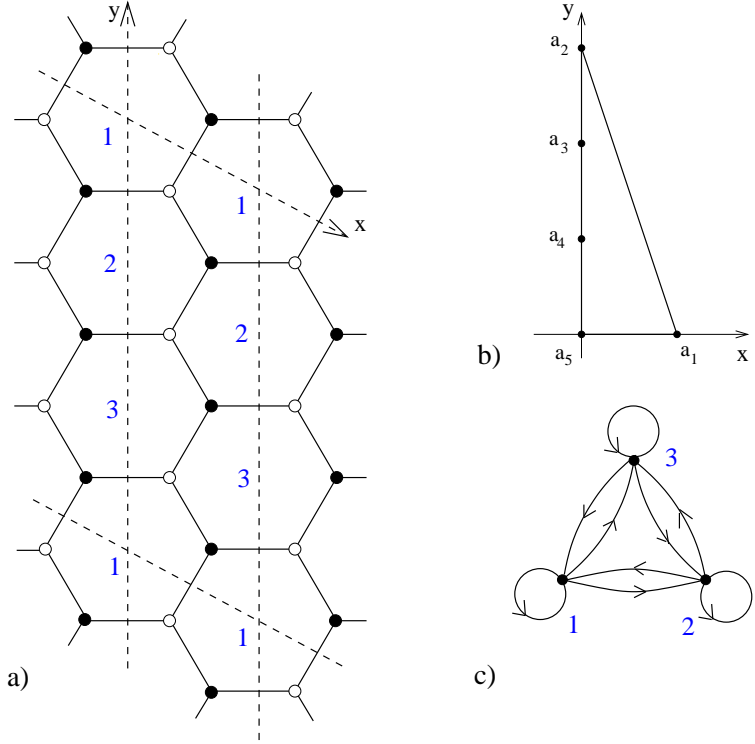} 
\caption{The $\mathbb C^3/\mathbb Z_3$ theory: a) dimer configuration b) toric diagram c) quiver diagram.}
\label{quotient}
\end{center}
\end{figure}

We can repeat this analysis for the whole family of generalized
conifolds of type $A_{k-1}$: they are obtained by modifying the toric
theories with geometry the quotient $\mathbb C^3/ \mathbb Z_k$. The
toric diagram $\mathcal{C}$ of $\mathbb C^3/ \mathbb Z_k$ is: 
\begin{equation}
(0,0,1) \quad (1,0,1) \quad (0,k,1)
\end{equation}
and its dual $\mathcal{C}^*$ is generated over the integers by the vectors:
\begin{equation}
(0,1,0) \quad (-k,-1,k) \quad (1,0,0) \quad (-1,0,1)
\label{csquot}
\end{equation}
The gauge theory has a dimer made up of $F=k$ gauge groups, $V=2k$
superpotential terms and $E=3k$ chiral fields. The faces are all
hexagons aligned in a single column and with the identifications as in
Figure \ref{quotient}a). Up to the adjoint fields the quiver is the
same as for the generalized conifolds of type $A_{k-1}$: among the
$3k$ fields there are $k$ $X_i$ and $k$ $\tilde X_i$, like for the
generalized conifolds, and the remaining fields are the adjoints
$Y_{ii}$, to which we will have to give mass. We draw in Figure
\ref{quotient} the dimer, toric diagram and quiver in the case $k=3$:
modifying this theory we will get again the generalized conifold of
type $A_2$. 

The minimal set of gauge invariants for the theory $\mathbb
C^3/\mathbb Z_k$ consists of $z$, $w$, $x_i$, $i=1,\ldots k$, defined
in (\ref{gaugeinvk}) plus the $k$ adjoints $Y_{ii}$. It is easy to see
that the $\Psi$-map for the toric theory is 
\begin{equation}
\begin{array}{l@{\quad}l}
w \rightarrow (0,1,0) & z \rightarrow (-k,-1,k) \\ 
Y_{11}, \ldots Y_{kk}\rightarrow (1,0,0) & x_1,\ldots x_k \rightarrow (-1,0,1)
\end{array}
\label{psimapquot}
\end{equation}
From this $\Psi$-map it is easy to see that the singularity is defined
by the equation: $z w=x_1^k$ in the four variables: $z$, $w$, $x_1$,
$Y_{11}$.  

The superpotential for the original toric theory is:
\begin{equation}
W_{T^3}=\sum_{i=1}^k Y_{ii}x_i-Y_{ii}x_{i-1}
\label{st3quot}
\end{equation}
where the index $i$ is periodic with period $k$. The massive terms
$Y_{ii}^2$ are mesons mapped to $(2,0,0)$, as it is obvious from
(\ref{psimapquot}), and again the quotient is with respect to
$\vec{d}=(2,0,0)-(0,0,1)=(2,0,-1)$. We have to add to the
superpotential (\ref{st3quot}) all mesons mapped to points: $(0,0,1)$,
$(2,0,0)$, $(-2,0,2)$. We find the superpotential: 
\begin{equation}
W_{T^2}=\left( \sum_{i=1}^k Y_{ii} \right) \left( \sum_{j=1}^k x_j \right)
+\left( \sum_{i=1}^k Y_{ii}^2 \right)+ \left( \sum_{i=1}^k x_i^2 \right)
\end{equation}
where again we have understood generic coefficients in front of every
term. Integrating out the massive fields $Y_{ii}$ one recovers the
superpotential of the generalized conifolds of type $A_{k-1}$
(\ref{supotk}), again with generic coefficients in front of each
term. Therefore we see that the $T^2$ theories of generalized
conifolds of type $A_{k-1}$ are obtained by quotienting the theories
$\mathbb C^3/\mathbb Z_k$ with respect to: $\vec{d}=(2,0,-1)$. 

We can perform again a change of coordinates using the same matrix $A$
in equation (\ref{aspp}), that sends $\vec{d}\rightarrow (0,0,1)$. The
vectors of $\mathcal{C}^*$ in (\ref{csquot}) are sent respectively to
the points: 
\begin{equation}
(1,0,0) \quad (k-1,k,-k) \quad (1,1,0) \quad (1,1,-1)
\end{equation}
and quotienting with respect to $(0,0,1)$ we find that the integer generators of $\mathcal{C}^*_{T^2}$ are:
\begin{equation}
w\rightarrow (1,0) \quad x_1,\cdots x_k \rightarrow (1,1) \quad z \rightarrow (k-1,k)
\label{csgen}
\end{equation}
Note that the points $(1,0)$, $(k-1,k)$ have multiplicities 1, whereas
the point $(1,1)$ has multiplicity 2, since, as discussed above, there
are $k-2$ independent linear relations between the mesons $x_1\ldots
x_k$, in agreement with the fact that there are 2 points in
$\mathcal{C}^*$ ($(1,1,0)$ and $(1,1,-1)$) projected to $(1,1)$. Note
that when $k=3$ the cone $\mathcal{C}^*_{T^2}$ is equivalent to that
in Figure \ref{sppres} b). 

It is not difficult to resolve the singularity $\mathbb C^3/\mathbb
Z_k$ and find through equation (\ref{ctoric}) \cite{MSY2} the
character $C_{T^3}(q)$; after changing coordinates with the matrix $A$
(\ref{aspp}) we obtain: 
\begin{equation}
C_{T^3}(q)=\frac{q_1 q_3}{\left( 1-q_1 q_2 \right) \left( q_3-q_1 q_2 \right)} 
\left[ \frac{1}{1-q_1}+ \frac{q_3^k}{q_3^k q_1-q_1^k q_2^k} \right]
\label{chquot}
\end{equation} 
The character $C_{T^2}(q)$ for the generalized conifold of type
$A_{k-1}$ is obtained by setting $q_3=1$ in the previous
formula. Performing the limit (\ref{limit}) on $C_{T^2}(q)$, or
equivalently on (\ref{chquot}) and then setting $b_3=0$, we find the
normalized volume for the basis of the generalized conifolds with a
Sasaki metric: 
\begin{equation}
V_{T^2}(b_1,b_3)=\frac{k}{b_1 \left( b_1+b_2 \right) \left( \left( k-1 \right) b_1 + k \, b_2\right)}
\label{volk}
\end{equation}
which in the case $k=3$ is equal to that in (\ref{volt2spp}) computed through the SPP.

The equations $\Delta=\vec{m}\cdot \vec{b}$ using the generating vectors in (\ref{csgen}) are:
\begin{equation}
b_1=\frac{3}{4}k \quad b_1+b_2=\frac{3}{2} \quad (k-1)b_1+k\, b_2=\frac{3}{4}k
\end{equation}
that allow to find the position of the Reeb vector for a
Sasaki-Einstein metric: $(b_1,b_2)=(3/4 \, k, 3/2-3/4\,k)$. Inserting
this value into (\ref{volk}) we find for the normalized volume of the
generalized conifold of type $A_{k-1}$: 
\begin{equation}
V_{T^2}=\frac{32}{27\, k}
\label{vollop}
\end{equation}
which again agrees with the results from a-maximization
(\ref{amaxak}), according to the AdS/CFT predictions
(\ref{matching}). 
Our formula for the volume (\ref{vollop}) agrees with the results of
reference \cite{Bergman}, which describes an alternative method for
computing $\textrm{Vol}(H)$ when the CY cone $C(H)$ is defined by a
single polynomial equation. 

\section{Other examples}
\label{examples}
In this Section we study the toric theories $Y^{2,1}$ and $L^{1,5;2,4}$
and try to modify them adding to their superpotential mesons mapped
to: $\vec{m}_0 + k \vec{d}$ in order to obtain theories with isometry
$T^2$.  
In particular our purpose is to see on concrete examples whether it is
possible to modify the toric theories keeping the moduli space
three-dimensional. Consider for instance the $\mathcal N=4$ quiver
gauge theory: it has three chiral fields and a superpotential with two
terms canceling in the abelianized version of the theory. If we add
superpotential terms that are non zero also when the gauge group is
$U(1)$ then we are introducing non trivial F-terms and hence the
dimension of the moduli space is necessarily less than three. A similar
situation happens in the conifold theory $T^{1,1}$: there are $4$
chiral fields, a superpotential equal to zero in the abelian case and
one D-term (recall that the number of independent D-terms is equal to
the number of gauge groups minus one: the sum of all charges for all
gauge groups is zero). Hence the moduli space is three dimensional,
but as soon as one introduces superpotential terms and hence non
trivial F-terms, the dimension of the moduli space is reduced. 

Therefore we choose here to start from the more complicated toric
theories $Y^{2,1}$ and $L^{1,5;2,4}$, that have more chiral fields and
a non zero superpotential also in the abelian theory. We add
superpotential terms as explained in Section \ref{mod} with all
generic coefficients (we suppose that the coefficients of the toric
terms are non zero), and try to see whether there are suitable choices
of these coefficients in the superpotential that allow the existence
of a three-dimensional moduli space.  

To do this we can work using the $E$ chiral fields as complex
variables: we write the F-term equations and solve for $E-F-2$ chiral
fields (supposing at generic points all the chiral fields different
from zero) in function of $F+2$ chiral fields, where $F$ is the number
of gauge groups. We have to impose that the remaining F-term equations
are satisfied: this gives non trivial constraints on the coefficients
in the superpotential. The D-terms (that are independent from the
superpotential) will reduce of $F-1$ the dimension of the $F+2$
dimensional manifold, leaving a three-dimensional cone. If the
conditions on the coefficients in the superpotential can be satisfied
also with non zero coefficients for (some of) the non toric terms,
then we have constructed three-dimensional non toric ($T^2$) complex
cones. An alternative and equivalent way is to work with mesons, we
associate a complex variable to each of the mesons that generate loops
in the quiver; the moduli space in this case can be expressed as a
(typically non complete) intersection in this space of complex
variables: since mesons are gauge invariants, we do not have to take
into account D-terms and quotients with respect to the corresponding
charges. There are non linear algebraic relations among mesons due to
their composition in terms of chiral fields, plus linear relations
induced by F-terms and depending on the coefficients appearing in the
superpotential. Imposing again that the resulting locus is three
dimensional we obtain the conditions on coefficients in the
superpotential. 

Consider now the theory for $Y^{2,1}$; we will not give all the details
that can be found in the literature \cite{benvenuti}. The toric
diagram $\mathcal{C}$ for $Y^{2,1}$ is generated by the vectors: 
\begin{equation}
A=(0,0,1) \quad B=(1,0,1) \quad C=(0,2,1) \quad D=(-1,1,1)
\label{cy21}
\end{equation}
The generators of the dual cone $\mathcal{C}^*$ over integer numbers are the nine vectors:
\begin{equation}
\begin{array}{lllll}
(0,1,0) & (-2,-1,2) & (1,-1,2) & (1,1,0) & (1,0,1) \\
(-1,0,1) & (-1,-1,2) & (0,-1,2) & (0,0,1) &
\end{array}
\label{csy21}
\end{equation}
In our analysis we have added to the superpotential mesons mapped to
these generators of $\mathcal{C}^*$, since they are irreducible mesons (they are
not the product of smaller loops) and hence simpler. For some choices
of the direction $\vec{d}$ along which to perform the quotient the
moduli space is a three dimensional $T^2$ cone; we sum up our results: 
\begin{equation}
\begin{array}{l@{\quad}l@{\quad}c}
\,\,\,\,\,\,\,\vec{d} & \textrm{mesons in superpotential} & 3d \, \textrm{complex cone} \, T^2\\
(1,0,0) & (1,0,1), (0,0,1), (-1,0,1) & \textrm{Yes} \\
(-1,-1,1) & (1,1,0), (0,0,1), (-1,-1,2) & \textrm{Yes} \\
(0,-1,1) & (0,1,0), (0,0,1), (0,-1,2) & \textrm{Yes} \\
(1,-1,1) & (0,0,1), (1,-1,2) & \textrm{No} \\
(-2,-1,1) & (0,0,1), (-2,-1,2) & \textrm{No}
\end{array}
\end{equation}

We have repeated the analysis for $L^{1,5;2,4}$ whose toric diagram $\mathcal{C}$ is:
\begin{equation}
(0,0,1) \quad (1,0,1) \quad (0,2,1) \quad (-2,1,1)
\end{equation}
The dual cone $\mathcal{C}^*$ is generated over integers by the vectors:
\begin{equation}
\begin{array}{llll}
(0,1,0) & (-2,-1,2) & (1,-2,4) & (1,2,0) \\
(-1,-1,2) & (-1,0,1) & (0,-1,2) & (0,0,1) \\
(1,-1,3) & (1,0,2) & (1,1,1) &
\end{array}
\end{equation}
Again we have added to the superpotentials mesons mapped to these generators of $\mathcal{C}^*$. The results are:
\begin{equation}
\begin{array}{l@{\quad}l@{\quad}c}
\,\,\,\,\,\,\,\vec{d} &  \textrm{mesons in superpotential} & 3d \, \textrm{complex cone} \, T^2\\
(0,-1,1) & (0,1,0), (0,0,1), (0,-1,2) & \textrm{Yes} \\
(-2,-1,1) & (0,0,1), (-2,-1,2) & \textrm{No} \\
(-1,-1,1) & (0,0,1), (-1,-1,2) & \textrm{Yes} \\
(-1,0,0) & (0,0,1), (-1,0,1) & \textrm{Yes} \\
(1,-2,3) & (0,0,1), (1,-2,4) & \textrm{No} \\
(1,-1,2) & (0,0,1), (1,-1,3) & \textrm{Yes} \\
(1,0,1) & (0,0,1), (1,0,2) & \textrm{Yes} \\
(1,1,0) & (0,0,1), (1,1,1) & \textrm{Yes} \\
(1,2,-1) & (0,0,1), (1,2,0) & \textrm{Yes} \\
\end{array}
\end{equation}

We see therefore that the existence of 3d complex cones (with isometry
group $T^2$) obtained by modifying toric theories is quite a common
feature. Indeed all the vectors $\vec{d}$ considered here along which
we performed the quotient satisfy the condition (\ref{finite}) for
having finite multiplicities (note that the proof of the equivalence
of a-maximization and volume minimization in the next Section relies
only on this property\footnote{we have also checked that all the field
theories obtained by quotienting with the vectors $\vec{d}$ considered
here satisfy the unitary bounds and have central charge $a \geq 1/4$; the
dual geometrical constraints have been recently considered in
\cite{GMSY}. Interestingly the field theory example considered in Section
(4.3) of the same work belongs to the general class of deformations
considered in this paper: with the notation of our Section \ref{lop},
take $k=2$, that is the toric theory $\mathbb{C}^3/\mathbb{Z}_2)$ and
take the quotient with respect to the
vector $\vec{d}=(p+1,0,-1)$. For $p>1$ it is easy to see that the
deformed gauge theory does not satisfy the unitary bound; the
matching between the central charge and the volume, computed in a
formal way, is satisfied because of the general mechanism of Section
\ref{proof}.}); yet for some of them there is no three
dimensional complex cone with reduced isometry $T^2$. It would be
interesting to understand whether there are conditions on $\vec{d}$
corresponding to the request of a three dimensional non toric moduli
space.

As a final remark we point out that it is not always obvious that the
deformation we add to the toric theory is relevant: in some cases
presented in this Section the new superpotential terms have dimension
greater than $3$ computed at conformal points of the original toric
theory. However recall that we are changing all the coefficients in the
superpotential (and possibly the gauge couplings) so that some
operator may become (marginally) relevant. We have checked on these
examples that the 
central charge $a$ of the deformed theories is always less than in
the original toric theories.

\begin{figure}
\begin{center}
\includegraphics[scale=0.6]{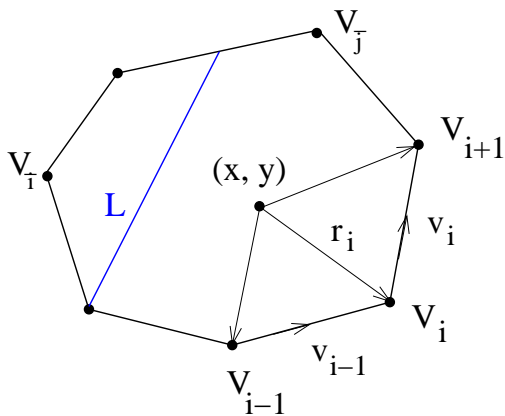} 
\caption{The toric diagram $P$ where volume minimization for the $T^3$
  theory must be performed. For $T^2$ theories volume minimization
  is performed on a line segment $L$ inside $P$.}
\label{toricdiagram}
\end{center}
\end{figure}
   
\section{a-maximization and volume minimization}
\label{proof}
In this Section we prove that the results of a-maximization and volume
minimization match the predictions of AdS/CFT (\ref{matching}) in the
general class of theories obtained by modifying toric theories as
explained in Section \ref{mod}. We make a proposal to identify the
space where volume minimization \cite{MSY2} should be performed when
the isometry is $T^2$ (in the case $T^1$ this locus is reduced to a
single point).
 
Let us start from a toric AdS/CFT correspondence: in our conventions
the vertices of the toric diagram are the points: $V_i=(x_i,y_i,1)$
that generate the fan $\mathcal{C}$. As in Section \ref{barmes} we also
introduce the sides of the toric diagram:
$v_i\equiv(x_{i+1},y_{i+1})-(x_i,y_i)$ and the vectors joining the
trial Reeb $(x,y)$ with the vertices: $r_i\equiv(x_i,y_i)-(x,y)$, see
Figure \ref{toricdiagram}. The Reeb vector for a Sasaki-Einstein
metric $\vec{b}=3(\bar{x},\bar{y},1)$ is the minimum of the volume
function for a Sasaki metric $V(\vec{b})$ when $\vec{b}$ lies on the
plane $z=3$: $\vec{b}=3(x,y,1)$, and the trial point $(x,y)$ lies
inside the toric diagram $P$ \cite{MSY} (the expression for the volume
of $H$ in function of $(x,y)$ is reported also in equation (\ref{volfun})). 
The proof of the equivalence
of volume minimization and a-maximization \cite{aZequiv} lies on the
fact that if we parametrize the charges $a_i$ as (\ref{aixy1}):
\begin{equation}
a_i \rightarrow a_i(x,y)\equiv \frac{2 l_i}{\displaystyle \sum_{j=1}^d
  l_j} \qquad \textrm{with} \qquad l_i\equiv\frac{\langle v_{i-1},
  v_{i} \rangle }{ \langle r_{i-1},v_{i-1} \rangle \langle r_i, v_i
  \rangle} 
\label{aixy}
\end{equation}  
then the matching (\ref{matching}) is true for every $(x,y)$ inside the toric diagram $P$ \cite{aZequiv}:
\begin{equation}
a(x,y)=\frac{1}{4 V(x,y)} \qquad \textrm{with} \qquad a(x,y)\equiv a(a_1(x,y),\ldots a_d (x,y))
\label{matcht3}
\end{equation}
where $V(x,y)$ is the normalized volume function $V(\vec{b})$ for a
Sasaki metric computed at the trial Reeb vector:
$\vec{b}=3(x,y,1)$. Note that parametrization (\ref{aixy})
automatically satisfies condition (\ref{erre}). Moreover the
derivatives of the trial charge $a(a_1,\ldots a_d)$ are zero along the
$d-3$ baryonic symmetries (\ref{baryonic}) and on the surface
$a_i=a_i(x,y)$ parametrized by $(x,y)$ \cite{aZequiv}: 
\begin{equation}
\sum_{i=1}^d B_i \left( \frac{\partial a}{\partial a_i} \right)_{|a_i \rightarrow a_i(x,y)}=0
 \qquad B_i \,\, \textrm{any baryonic symmetry:} \,\, \sum_{i=1}^d B_i \vec{V}_i=0
\label{barzero} 
\end{equation}
Therefore also a-maximization in the $d-1$ independent parameters
$a_i$ is reduced to an extremization problem in the variables $(x,y)$.  

Moreover, as proved in \cite{MSY,aZequiv}, see also Section \ref{barmes}, 
for every $(x,y)$ inside the toric diagram $P$ we have the following relations:
\begin{equation}
\sum_{i=1}^d a_i(x,y) \vec{V}_i=2(x,y,1) \quad \Rightarrow \quad 
\frac{3}{2}\sum_{i=1}^d \left( \vec{m}\cdot \vec{V}_i  \right) a_i(x,y)= 3(x,y,1)\cdot \vec{m}
\label{same}
\end{equation}
where the second equality is an extension of
$\Delta=\vec{b}\cdot\vec{m}$ to the whole interior of the polygon
$P$. $\vec{m}$ is a point in $\mathcal{C}^*$ and as usual represents the charges
of a meson under $\Psi$-map.

Let us now modify the original toric $T^3$ theory to a $T^2$ theory by
adding superpotential terms of the form $\vec{m}_0+k \vec{d}$, as in
Section \ref{mod}; we will suppose that this gauge theory (for some
suitable choice of coefficients in the superpotential) admits a
three-dimensional moduli space of vacua and that multiplicities of
holomorphic functions over this three-dimensional cone are counted as
explained in Section \ref{mod}. The first step is to try to understand
a-maximization: the trial $a$ charge is the same function
$a(a_1,\ldots a_d)$ of the toric case but we have to add the new
constraint (\ref{new}) besides the usual condition
(\ref{erre}). Consider the plane $Q$ perpendicular to the vector
$\vec{d}$ and passing through the origin. The plane containing the
toric diagram $P$ is perpendicular to $\vec{m}_0\equiv(0,0,1)$: it is
the plane $z=1$ in our conventions. These two planes in $\mathbb R^3$
intersect on a line since $\vec{d}$ cannot be proportional to
$\vec{m}_0\in \mathcal{C}^*$ because of condition (\ref{finite}) and in
particular the plane $Q$ intersects the polygon $P$ on a line segment
$L$ interior to $P$, look at Figure \ref{toricdiagram}; in fact
condition (\ref{finite}) is equivalent to the request that there exist
$\bar i$, $\bar j$ such that $\vec{d}\cdot \vec{V}_{\bar i}>0$ and
$\vec{d}\cdot \vec{V}_{\bar j}<0$. The points $(x,y)\in L$ are
characterized by: $\vec{d}\cdot (x,y,1)=0$.

Interestingly if we restrict $(x,y)$ to vary along the line $L$ and
use the same parametrization (\ref{aixy}) and equation (\ref{same}) we
find that the new request (\ref{new}) in the case $T^2$ is satisfied: 
\begin{equation}
\sum_{i=1}^d \left( \vec{d} \cdot \vec{V}_i \right) a_i(x,y)=0 \qquad
\quad \forall (x,y)\in L 
\end{equation}
a-maximization must be performed on $d-2$ independent variables $a_i$;
but note that since $L\subset P$ and since, as observed in Section
\ref{mod}, the $d-3$ baryonic symmetries $B_i$ are the same for the
$T^3$ and the $T^2$ (or $T^1$) theories then equation (\ref{barzero})
tells us that along $L$ we have decoupled baryonic symmetries also for
the $T^2$ theory. a-maximization is reduced to an extremization
problem in one variable, as expected since there is now only one
flavor symmetry. We have to maximize the same function $a(x,y)$ of the
toric case but along $L$ and not on the whole interior of $P$. Note
therefore that $a_{T^2}\leq a_{T^3}$ and recall that in our theories
(that have $\textrm{tr}R=0$ and hopefully a supergravity dual) the
central charges $a$ and $c$ are equal. This is in agreement with the
decreasing of $a=c$ along the renormalization group flow: we have added
relevant terms to the $T^3$ theory and then flown to new
superconformal IR fixed points, corresponding to $T^2$ theories.

Let us consider now meson charges; after the introduction of the
superpotential terms that break one flavor symmetries, aligned vectors
$\vec{m} +k \vec{d}$ in the toric $\mathcal{C}^*$ are identified: we have to
project perpendicularly each integer point $\vec{m}\in \mathcal{C}^*$ on the
plane $Q$ to find the new cone of charges for the theory $T^2$. The
scaling dimensions of mesons in the $T^2$ theory can be computed as
usual with the $\Psi$-map: $\Delta=3/2 \sum_i (\vec{m} \cdot
\vec{V}_i) a_i(\bar x, \bar y)$ and we know from the previous analysis
that $(\bar x,\bar y)$ lie on $L$. The formula for the scaling
dimensions of mesons can be extended to the equality, which holds for
all $(x,y)\in L$: 
\begin{equation} 
\frac{3}{2}\sum_{i=1}^d \left( \vec{m}\cdot \vec{V}_i  \right) a_i(x,y)= 3(x,y,1)\cdot \vec{m} \qquad
\vec{m}\sim \vec{m} + k \vec{d}, \quad (x,y) \in L
\label{deltat2}
\end{equation}
which is the same as (\ref{same}), but we underline that both sides of
the equality are invariant under $\vec{m}\sim \vec{m} + k \vec{d}$ if
$(x,y)\in L$, again because of (\ref{same}). All of this tells us that
the cone of charges for holomorphic functions of the $T^2$ theory is
obtained by projecting perpendicularly each integer point $\vec{m}$ in
the toric $\mathcal{C}^*$ on the plane $Q$; we will call $\pi$ the orthogonal
projection over $Q$ (\ref{pi}). The Reeb vector $\vec{b}$ lies now on
$Q$ and the scalar product in $\Delta=\vec{b}\cdot \vec{m}$ is the
restriction of the scalar product of the toric theory to the plane
$Q$.  
In our examples we computed the Reeb vector from a-maximization
imposing that for all mesons $\Delta=\vec{b}\cdot \vec{m}$ and it is
evident from this discussion that these equations are always
consistent since they admit the solution $\vec{b}=3(\bar x, \bar y,
1)$, with $(\bar x,\bar y)\in L$ coming from a-maximization $a(x,y)$
along $L$. 

Let us now try to understand the volume function $V_{T^2}(\vec{b})$
for a Sasaki metric on the base of the cone, where $\vec{b}$ varies
inside $Q$. Suppose to perform an $SL(3,\mathbb Z)$ that sends
$\vec{d}$ in, say, $(1,0,0)$; this is always possible since $\vec{d}$
is a primitive integer vector. In this frame the plane $Q$ is the
plane $(y,z)$. Under the hypothesis that multiplicities of holomorphic
functions are computed, as explained in Section \ref{mod}, by counting
the number of integer points of $\mathcal{C}^*$ projected to the same point on
the plane $Q$, then the character $C_{T^2}(q_2,q_3)$ is equal to
$C_{T^3}(1,q_2,q_3)$. And from equation (\ref{limit}) it follows
easily: 
\begin{equation}
V_{T^2}(b_2,b_3)=\lim_{t \rightarrow 0} t^3 C_{T^2}(e^{-t b_2},e^{-t b_3})=
\lim_{t \rightarrow 0} t^3 C_{T^3}(1,e^{-t b_2},e^{-t b_3})=V_{T^3}(0,b_2,b_3)
\end{equation}
More generally we see that the function $V_{T^2}$ is simply the volume
function of the toric case $V_{T^3}$ restricted to the plane $Q$
\footnote{Note that since we have an expression of charges $a_i$ in
  function only of the Reeb vector (\ref{aixy}) and because of the AdS/CFT
  prediction (\ref{bardim}), one may conjecture that also the volume
  functions of divisors for Sasaki, not necessarily Einstein, metrics
  may be obtained in the $T^2$ theory by restricting to the plane $Q$
  the volume functions $\textrm{Vol}_{\Sigma_i} (b)$ of
  the original toric theory. Of course this should be the case at
  least in the point of volume minimization, according to AdS/CFT. We
  are grateful to C.~P.~Herzog for useful discussions on this point.}. 

\begin{figure}
\begin{center}
\includegraphics[scale=0.55]{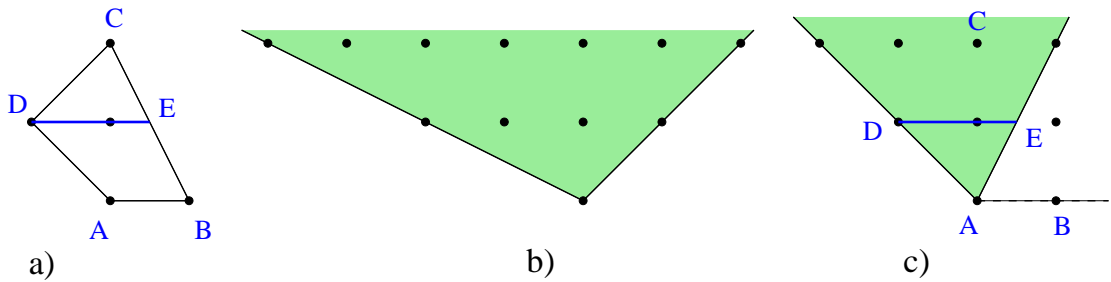} 
\caption{a) Toric diagram for $Y^{21}$ with the line $L\equiv D\,E$
  corresponding to the quotient $\vec{d}=(0,-1,1)$. 
b) The cone $\mathcal{C}^*_{T^2}$ of charges for holomorphic
functions. c) The direct cone $\mathcal{C}_{T^2}$ in green, dual of
$\mathcal{C}^*_{T^2}$, where the function $V_{T^2}$ is defined.} 
\label{coney21}
\end{center}
\end{figure}

It is clear at this point that the matching $a_{T^2}=1/(4 V_{T^2})$
always holds with $V_{T^2}$ equal to the volume function computed at
the Reeb vector $\vec b$ coming from a-maximization; more generally,
as a simple restriction of (\ref{matcht3}), the two trial functions
$a_{T^2}(x,y)$ and $V_{T^2}(x,y)$ agree on the whole line segment $L$: 
\begin{equation}
a_{T^2}(x,y)=\frac{1}{4 V_{T^2}(x,y)} \qquad (x,y) \in L
\end{equation}
It is natural at this point to identify the line $L$ as the locus
where to perform volume minimization \cite{MSY2}. We suggest that this
is the case for the class of non toric theories introduced in Section
\ref{mod}. 

As an example consider the toric theory $Y^{21}$ with $\mathcal{C}$
and $\mathcal{C}^*$ generated by vectors in (\ref{cy21}) and
(\ref{csy21}) respectively, and consider the $T^2$ quotient with
vector $\vec{d}=(0,-1,1)$, that admits a three dimensional cone as a
moduli space of vacua, as told in Section \ref{examples}. Line $L$ is
the intersection of the plane $Q$ perpendicular to $\vec{d}$:
$-y+z=0$, with the plane of the toric diagram: $z=1$, and hence is the
line segment with $z=1$, $y=1$ with extremal points $D=(-1,1,1)$ and
$E=(1/2,1,1)$; it is drawn in Figure \ref{coney21}a). 
We can apply the $SL(3,\mathbb Z)$ transformation
$S=((1,0,0),(0,1,1),(0,0,1))$ to the points of $\mathcal{C}^*$; the
vector $\vec{d}$ is mapped to $(0,0,1)$. The first four generating
vectors in (\ref{csy21}) that generate $\mathcal{C}^*$ over real
numbers are mapped by $S$ into: 
\begin{equation}
(0,1,0) \quad (-2,1,2) \quad (1,1,2) \quad (1,1,0)
\end{equation}
With these coordinates the cone
$\mathcal{C}^*_{T^2}=\pi(\mathcal{C}^*)$ is obtained by projecting
these vectors on the plane $Q=(x,y)$ ($\pi$ stands for the orthogonal
projection on the plane $Q$). Hence $\mathcal{C}^*_{T^2}$ is generated
over real numbers by $(-2,1)$ and $(1,1)$ and it is drawn in Figure
\ref{coney21}b). Transforming the points in (\ref{cy21}) with $^tS^{-1}$
we get: 
\begin{equation}
A=(0,0,1) \quad B=(1,0,1) \quad C=(0,2,-1) \quad D=(-1,1,0)
\end{equation}
and the point $E$ in the new coordinates is $E=(1/2,1,0)$.
In Figure \ref{coney21} we show the projection of $\pi(\mathcal{C})$
on the plane $Q$; it is generated by $(1,0)$, $(-1,1)$ over real
numbers. In the same Figure we also draw in green the cone over $L$,
that in the intersection $\mathcal{C} \cap Q$, generated by vectors
$(1,2)$, $(-1,1)$: on the interior of this two dimensional cone the
function $V_{T^2}$ is positive and well defined, and it is convex on
the segment $L=D\,E$ where volume minimization has to be performed. 
Interestingly we note that the cone $\mathcal{C} \cap Q$ is just the
dual of $\mathcal{C}^*_{T^2}$, and hence we will call it
$\mathcal{C}_{T^2}$\footnote{note that $\mathcal{C}_{T^2} \subseteq
  \pi(\mathcal{C})$, but they are generally different, as in the
  example of Figure \ref{coney21} c).}: 
\begin{equation}
\mathcal{C}_{T^2} \equiv \left( \mathcal{C}^*_{T^2} \right)^* \equiv \left( \pi(\mathcal{C}^*) \right)^* = \mathcal{C} \cap Q
\label{theorem}
\end{equation}  
To see (\ref{theorem}) suppose that $\vec{n}\in \mathcal{C}\cap Q$,
then $\vec{n} \cdot \vec{m}\geq 0$, for any $\vec{m}\in
\mathcal{C}^*$. But since $\vec{n}$ belongs to $Q$, then $\vec{n}
\cdot \vec{m}=\vec{n} \cdot (\vec{m}+k \vec{d}) \geq 0$ and we can
choose $k$ to obtain: $\vec{n} \cdot \pi(\vec{m})\geq 0$, hence we
have shown that: $\mathcal{C} \cap Q \subseteq
\mathcal{C}_{T^2}$. Conversely if $\vec{n}$ is in $\mathcal{C}_{T^2}$
then it is a vector in $Q$ such that $\vec{n} \cdot \pi(\vec{m}) \geq
0$ for any $\vec{m}\in \mathcal{C}^*$. But again for vectors
$\vec{n}\in Q$ we have: $\vec{n} \cdot \vec{m}=\vec{n}\cdot
\pi(\vec{m})\geq 0$ for any $\vec{m}\in \mathcal{C}^*$, therefore
$\vec{n}$ also belongs to $\mathcal{C}$. Hence also $\mathcal{C}_{T^2}
\subseteq \mathcal{C} \cap Q$ holds. 

We can repeat the same analysis for $T^1$ theories: there are now two
independent vectors $\vec{d}_1$, $\vec{d}_2$, and two further
constraints to be imposed on trial charges $a_i$ (\ref{newt1}). Let
$Q_1$ and $Q_2$ be the planes perpendicular to $\vec{d}_1$ and
$\vec{d}_2$ respectively, and $L_1$ and $L_2$ their intersection with
the toric diagram $P$ on the plane $z=1$. Then the two constraints
(\ref{newt1}) are satisfied by $a_i(x,y)$ in (\ref{aixy}) on the point
$N$ defined as the intersection of the line segments $L_1$ and
$L_2$. Note that $N$ lies in the interior of the convex polygon $P$;
in fact the point $N$ is the intersection of the plane $z=1$ with the
line $R$ passing through the origin and perpendicular to the plane
generated by $\vec{d}_1$, $\vec{d}_2$. This line $R$ belongs to the
direct cone $\mathcal{C} \cup (-\mathcal{C})$ as explained in Section
\ref{mod}. a-maximization must be performed on a space of $d-3$
independent $a_i$, and at point $N$ there are just $d-3$ baryonic
directions with zero derivatives for the trial $a$ function. The Reeb
vector is determined in terms of the coordinates of $N$ as:
$\vec{b}=3(x_N,y_N,1)$ and lies on $R$. The cone for mesonic charges
can be obtained as the projection of the toric $\mathcal{C}^*$ on the
line $R$; we will call $\Pi$ the orthogonal projection over the line
$R$ (\ref{Pi}). Equality (\ref{deltat2}) is still true and non
ambiguous under $\vec{m}\sim \vec{m}+k \vec{d}_1 +h \vec{d}_2$ if
$(x,y)$ coincides with $N$. The volume function $V_{T^1}(\vec{b})$ is
a restriction of $V_{T^3}(\vec{b})$ to the line $R$. There is no
volume minimization to perform; the central charge $a$ and the volume
of a Sasaki-Einstein metric agree (\ref{matching}) again as a simple
restriction of the identity (\ref{matcht3}) to the point $N$.

\vspace{2em}
\noindent {\Large{\bf Acknowledgments}}

\vspace{0.5em}

We would like to thank Sergio Benvenuti, Amihay Hanany, Christopher
P.~Herzog, Dario Martelli, David Vegh and in particular Alessandro
Tomasiello for invaluable
discussions. D.F. is grateful to Elena Andreini, Giulio Bonelli,
Barbara Fantechi, Luca Martucci for very useful and patient
discussions, to Angel M.Uranga for
helpful correspondence and to Andrea Brini, Diego Gallego and
Houman Safaai for constant support. We thank MIT for kind
hospitality while part 
of this work was done. This work is supported in part by INFN (and
Bruno Rossi exchange program) and MURST under  
contract 2005-024045-004  and 2005-023102 and by 
the European Community's Human Potential Program
MRTN-CT-2004-005104. 

\vspace{1em}

\appendix
 
\section{Review of toric geometry and quiver gauge theory}
\label{dimtor}
The correspondence between quiver gauge theories (considering equivalent two
theories that flow to the same IR fixed point) and singularities is complete
in the toric case \cite{dimers,rhombi,mirror}; we will introduce now 
the elements of this correspondence that will be used in the following.
Roughly speaking a six dimensional manifold is toric 
if it has at least $U(1)^3$ isometry. All geometrical information about 
toric CY cones are encoded in the fan  $\mathcal{C}$ \cite{fulton}, 
in this case a cone in $\mathbb{Z}^3$ defined by $d$ vectors $V_i$. 
The Calabi-Yau condition requires that all vectors $V_i$ lie 
on a plane. In our conventions the vectors $V_i$ will lie on the plane with 
third coordinate $z=1$. By ignoring the third coordinate, we can
consider the \emph{toric diagram}, a convex polygon in the plane with integer
vertices, which is just the
intersection of the fan with the plane $z=1$. 
The \emph{(p,q) web} is the set of vectors perpendicular to the edges of the 
toric diagram and with the same length as the corresponding edges.
In the toric case the gauge theory is completely identified by the
\emph{periodic quiver}, a diagram drawn on $T^2$ (it is the ``lift'' 
of the usual
quiver to the torus): nodes represent $SU(N)$ gauge groups, oriented
links represent chiral bifundamental multiplets and faces represent
the superpotential: the trace of the
product of chiral fields of a face gives a superpotential 
term. 
Equivalently the gauge theory is described by the
\emph{dimer configuration}, or \emph{brane tiling},
the dual graph of the periodic quiver, drawn also on a torus $T^2$.
In the dimer the role of faces and vertices is exchanged: 
faces are gauge groups and vertices are superpotential terms.
The dimer is a bipartite graph: it has an equal number of white and 
black vertices (superpotential terms with sign + or - respectively)  
and links connect only vertices of different colors.

By applying Seiberg dualities to a quiver gauge theory we can obtain
different quivers that flow in the IR to the same CFT: to a toric
diagram we can associate different quivers/dimers describing the same
physics. It turns out that in the toric case one can always find phases where all the 
gauge groups have the same number of colors; these are called 
\emph{toric phases}. Seiberg dualities keep constant the number of
gauge groups $F$, but may change the number of fields $E$, and
therefore the number of superpotential terms $V=E-F$ (since the dimer is on a torus we have: $V-E+F=0$.). 
A zig-zag path in the dimer is a path of links that turn maximally 
left at a node, maximally right at the next node, then again 
maximally left and so on \cite{rhombi}. 
It was noted that every link of the dimer belongs to exactly 
two different zig-zag paths, oriented in opposite directions. 
Moreover for dimers representing consistent theories the zig-zag paths 
are closed non-intersecting loops. There is a \emph{one to one
correspondence between zig-zag paths and legs of the $(p,q)$ web}: 
the homotopy class in the fundamental group of the torus of every 
zig-zag path is given by the integer numbers $(p,q)$ of the 
corresponding leg in the $(p,q)$ web and this fact was used in
\cite{rhombi} to find a general algorithm to reconstruct the gauge
theory from the geometry in the toric case. 

Non anomalous $U(1)$ symmetries play a very important role in the
gauge theory. Here we review how to count and parametrize them and
how to compute the charge of a certain link in the dimer.
For smooth horizons $H$ we expect $d-1$ global non anomalous symmetries, 
where $d$ is the number of sides of the toric diagram in the dual theory.
We can count these symmetries from the 
number of massless vectors in the $AdS$ dual. Since the manifold is toric, 
the metric has three $U(1)$ isometries.
One of these (generated by the Reeb vector) corresponds to the
R-symmetry while the other two give two global flavor symmetries 
in the gauge theory. Other gauge fields in $AdS$ come 
from the reduction of the RR four form on the non-trivial three-cycles
in the horizon manifold $H$, and there are $d-3$ three-cycles in
homology \cite{tomorrow}  when $H$ is smooth.
On the field theory side, these gauge fields  correspond to baryonic 
symmetries.
Summarizing, the global non anomalous symmetries are:
\begin{equation}
U(1)^{d-1}=U(1)^2_F \times U(1)^{d-3}_B
\label{count}
\end{equation}
If the horizon $H$ is not smooth (that is the toric diagram has integer
points lying on the edges), equation (\ref{count}) is still true with $d$
equal to the perimeter of the toric diagram in the sense of toric
geometry (d = number of vertices of toric diagram + number of integer
points along edges).
In this paper we use the fact that these $d-1$ global non anomalous charges can be parametrized by $d$
parameters $a_1, a_2, \ldots ,a_d$ \cite{aZequiv}\footnote{The
algorithm proposed in \cite{aZequiv} to extract the field theory
content from the toric diagram is a generalization of previously known
results, see for instance \cite{benvenuti,tomorrow,hananymirror}, 
and in particular of the folded quiver in \cite{kru2}.}, each associated 
with a vertex of the toric diagram or a point along an edge, satisfying the constraint:
\begin{equation}
\sum_{i=1}^d a_i = 0
\label{sum}
\end{equation}
The $d-3$ baryonic charges are those satisfying the further
constraints \cite{tomorrow}:
\begin{equation}
\sum_{i=1}^d a_i V_i=0
\label{bar}
\end{equation}
where $V_i$ are the vectors of the fan: $V_i=(x_i,y_i,1)$ with
$(x_i,y_i)$ the coordinates of integer points along the perimeter of
the toric diagram. 
The R-symmetries are parametrized with the $a_i$
in a similar way of the other non-baryonic global symmetry, but they will satisfy the different constraint
\begin{equation}
\sum_{i=1}^d a_i = 2
\label{sumr}
\end{equation} 
due to the fact that the terms in the superpotential must have
$R$-charges equal to two.

\begin{figure}[h!]
\centering
\includegraphics[scale=0.5]{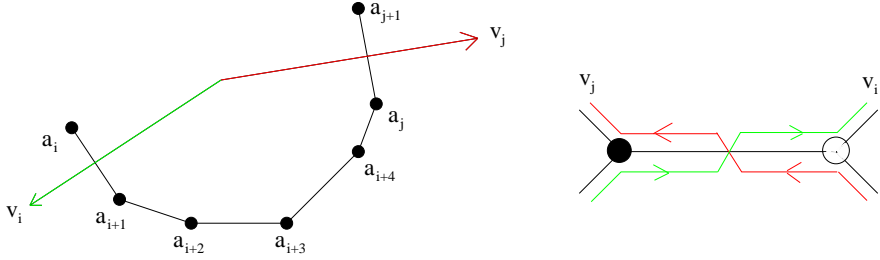}
\caption{(i) On the left: part of a toric diagram $P$ with the charge
  distribution (a trial charge $a_i$ for every vertex $V_i$ ) and two
  vectors $v_i$ and $v_j$ of the $(p,q)$-web. (ii) On the right:
  the corresponding zig-zag paths and the link with charge $a_{i+1} +
  a_{i+2} + a_{i+3}+a_{i+4}+a_j$ according to the recipe explained in
  the main text.}\label{ziza} 
\end{figure}

The simplest algorithm to compute the charge of a
generic link in the dimer in function of the parameters $a_i$ is probably the one that make use of the zig-zag paths.
Consider the two zig-zag paths to which a link in the dimer belongs. 
They correspond to two vectors $v_i=(p_i, q_i)$ and $v_j=(p_j, q_j)$ 
in the $(p,q)$ web. Then the charge of the link is given by the sum 
of the parameters $a_{i+1}+a_{i+2} \ldots +a_{j}$ between the vectors
$v_i$ and $v_j$ as shown in Figure \ref{ziza}.  
This rule explains the formula for the multiplicities of fields 
with a given charge \cite{aZequiv}: 
since every link in the dimer corresponds to the intersection of two
zig-zag paths, the number of fields with charge $a_{i+1}+a_{i+2} \ldots +a_{j}$
is equal\footnote{This is true in minimal toric phases (the toric phases with the minimal number of fields ), where the
  number of real intersections between two zig-zag paths is equal to
  the topological number of intersections.} to the number of 
intersections between the zig zag paths 
corresponding to $v_i$ and $v_j$, which is just $\mathrm{det}(v_i,v_j)$.

One can show that this parametrization automatically
solve the conditions of vanishing of gauge and superpotential beta functions 
\cite{aZequiv,proc}. The $d-1$ independent free quantities $a_i$ parametrize
the $d-1$ global abelian symmetry of the gauge theory that can mix with the
R-symmetry. The value of $a_i$ at the fixed point can be found by using 
a-maximization.

\section{The moduli spaces for $(P)dP_4$ theories}
We give here more details about the moduli spaces of the $(P)dP_4$
theories considered in Section \ref{pdp4ex}. 
As usual for toric theories, the moduli space of the $T^3$ $PdP_4$ is
determined by the linear relations among the generating vectors of the
cone $\mathcal{C}^*$. We associate with every generator of
$\mathcal{C}^*$ in (\ref{gen}) a complex coordinate $z_1$, $z_2$,
$z_3$, $z_4$, $z_5$ and $t$ and choose a representative meson among
those (F-term equivalent) mapped to the point of $\mathcal{C}^*$: 
\begin{equation}
\begin{array}{c}
n_4 \rightarrow (0,-1,2) \rightarrow z_4 \quad q_1 \rightarrow (-1,-1,3) \rightarrow z_3 
\quad n_8 \rightarrow (0,1,0) \rightarrow z_1 \\
m_6 \rightarrow (0,0,1) \rightarrow t \quad n_1 \rightarrow (-1,0,2) \rightarrow z_2 \\
n_5 \rightarrow (1,0,0) \rightarrow z_5
\label{name}
\end{array}
\end{equation}
where for the complex coordinates we have used the same notation as in \cite{ago}.
The equations defining the toric manifold are therefore the five quadrics in $\mathbb C^6$:
\begin{equation}
z_1 z_3=z_2 t \quad z_2 z_4 =z_3 t \quad z_3 z_5=z_4 t \quad z_2 z_5=t^2 \quad z_1 z_4 =t^2
\label{modulit3}
\end{equation}
This is a non complete intersection and it is easy to verify that the
moduli space is three-dimensional. 

The moduli spaces in the cases of the theories $T^2$ and $T^1$ are
more complicated; interestingly they are still defined as a non
complete intersection of five quadrics in $\mathbb C^6$, with
equations generalizing those in (\ref{modulit3}); see also
\cite{wijnholt} for the $T^1$ theory describing $dP_4$.  
In the non toric case $T^2$ there may be also a deformation of
the complex structure that leaves the manifold a cone. 
We will deal in this Appendix with the easier case of the $T^2$
$PdP_4$, with the general superpotential: 
\begin{equation}
W_{T^2}= a\,(m_1+m_3+m_5+m_7)-b\,(m_2+m_4+m_6+m_8)+c\, n_1 -d\, n_3+ e\, n_5 -f\, n_7 
+g \,p_1 
\label{supert2}
\end{equation} 
where it is clear that we can rescale the coefficients of the toric
terms in this way (if they are non zero). It is easy to see that there
are suitable choices of the coefficients in (\ref{supert2}) allowing a
three dimensional moduli space: there are 15 fields and 7 gauge
groups, hence 6 independent D-terms (D-terms equations and quotients
are independent from the parameters in the superpotential). Therefore
if we want a three dimensional manifold, F-terms should reduce the
dimension of the manifold of 6: we solve some of the F-term equations
with respect to 6 fields (supposing at generic points all chiral
fields different from zero); if we impose that the remaining F-term
equations are satisfied with these substitutions, we get some
complicated conditions on the coefficients in (\ref{supert2}). It is
not difficult to see that these conditions are satisfied with the
choice: 
\begin{equation}   
e=\frac{b^2 f}{a^2-d\, f} \quad 
g=\frac{a\, b^4-a^5+2 a^3 d \,f-a \, d^2 f^2}{b^3 f} \quad
c=\frac{a^4-b^4-a^2 d \,f}{b^2 f}
\label{3dcond}
\end{equation}
Note that there are other branches of solutions that assure a
three-dimensional complex moduli space; in the following we will refer
to (\ref{3dcond}). The same conditions could have been obtained using
directly gauge invariant mesons. 

To write the moduli space using only algebraic equations we have to
consider the relations among mesons. There are non linear relations
following directly from the definitions (\ref{mesdef}) in terms of
elementary fields of the 24 generating mesons; hence these relations
do not depend on the superpotential but only on the quiver. In our
example these non linear relations define a complex 9 dimensional
submanifold of $\mathbb C^{24}$. A peculiar role is played by the
first 20 mesons in (\ref{mesdef}), that is those mapped to generating
vectors of the toric $\mathcal{C}^*$; the quadratic relations among
them are: 
\begin{equation}
\begin{array}{lllll}
m_1 m_4=n_7 p_1 & m_1 m_5= n_4 n_6 & m_1 m_7 =n_3 n_7 & m_1 n_1=m_8 p_1 & m_1 n_2=n_7 q_1 \\ 
n_4 p_1=m_1 q_2 & m_2 m_3=n_8 p_2 & m_2 m_6 = n_3 n_5 & m_2 m_8 = n_4 n_8 & m_2 n_1 = n_8 q_2 \\
m_2 n_2 = m_7 p_2 & n_3 p_2 = m_2 q_1 & m_3 m_5 = n_1 n_5 & m_3 m_7= n_2 n_8 & m_3 n_3 =n_8 q_1 \\
m_3 n_4 = m_8 p_2 & n_1 p_2 = m_3 q_2 & m_4 m_6 = n_2 n_6 & m_4 m_8 = n_1 n_7 & m_4 n_3 = m_7 p_1 \\ 
m_4 n_4 = n_7 q_2 & n_2 p_1 = m_4 q_1 & m_5 p_1 = n_6 q_2 & m_5 p_2 = n_5 q_2 & m_5 q_1 = m_6 q_2 \\
m_6 p_1 = n_6 q_1 & m_6 p_2 = n_5 q_1 & n_2 n_3 = m_7 q_1 & n_1 n_4 = m_8 q_2 & n_5 p_1 = n_6 p_2
\end{array}
\label{quad}
\end{equation} 

There are no cubic independent relations and a single independent
quartic equation among the first 20 mesons in (\ref{mesdef}): $m_5 m_6
n_7 n_8=m_7 m_8 n_5 n_6$. 

Then we have to intersect this 9 dimensional manifold with the linear
space following from F-term relations (\ref{lineart21}),
(\ref{lineart22}) and (\ref{lineart23}). Moreover it is easy to see
that, once F-term relations are taken into account, the mesons $t_1$,
$t_2$, $t_3$ and $t_4$ are equal to quadratic polynomials in the first
20 mesons in (\ref{mesdef}), and therefore they can be eliminated. In
fact it is easy to check that also all non linear relations involving
the $t_i$ do not add other independent equations to those we will
write. Let us consider therefore the first 20 mesons in
(\ref{mesdef}): we can solve the linear constraints (\ref{lineart21}),
(\ref{lineart22}) and (\ref{lineart23}) expressing mesons in function
of the 6 mesons: $n_4$, $q_1$, $n_8$, $m_6$, $n_1$,
$n_5$. Substituting these linear relations in (\ref{quad}), using
conditions (\ref{3dcond}), we see that there remain only 5 linearly
independent quadratic equations: 
\begin{eqnarray}
& & \left( a^2 - d\,f \right) \,\left( -\left( b^4\,t\,z_2 \right)  + 
     a^2\,\left( a^2 - d\,f \right) \,t\,z_2 + 
     b^3\,f\,\left( t^2 - z_1\,z_4 \right)  \right)  - b^4\,f^2\,t\,z_5  = 0 \nonumber \\[0.5em]
& & \left( \left( a^2 - d\,f \right) \,
     \left( b^3\,f\,t^2 + a^2\,\left( a^2 - d\,f \right) \,t\,z_2 - 
       a\,b\,\left( a^2 - d\,f \right) \,z_1\,z_3 \right)  \right)  -
  b^4\,f^2\,t\,z_5  =  0  \nonumber \\[0.5em]
& & \left( b^2\,{\left( a^2 - d\,f \right) }^2\,t^2 \right)  - 
  \left( a^2 - d\,f \right) \,\left( 2\,b^3\,f\,t + 
     a^2\,\left( a^2 - d\,f \right) \,z_2 \right) \,z_5 + b^4\,f^2\,{z_5}^2  = 0  \nonumber \\[0.5em] 
& & a\,b^5\,d^2\,f^2\,z_2\,z_3 - a^9\,\left( f\,t + b\,z_2 \right) \,z_3 + 
  3\,a^7\,d\,f\,\left( f\,t + b\,z_2 \right) \,z_3  \nonumber \\ & & + 
  a^5\,\left( b^5\,z_2 - 3\,d^2\,f^2\,\left( f\,t + b\,z_2 \right)  \right)
     \,z_3 + a^3\,d\,f\,\left( -2\,b^5\,z_2 + 
     d^2\,f^2\,\left( f\,t + b\,z_2 \right)  \right) \,z_3  \nonumber \\ & & + 
  a^6\,b^2\,f\,\left( f\,t + b\,z_2 \right) \,z_4 - 
  2\,a^4\,b^2\,d\,f^2\,\left( f\,t + b\,z_2 \right) \,z_4  \nonumber \\ & & + 
  a^2\,b^2\,f^2\,\left( b^4\,t + d^2\,f^2\,t + b\,d^2\,f\,z_2 \right) \,
   z_4 - b^6\,f^3\,z_4\,\left( d\,t + b\,z_5 \right) = 0  \nonumber \\[0.5em] 
& & \left( \left( a^2 - d\,f \right) \,t\,
     \left( -\left( a^4\,z_3 \right)  + b^4\,z_3 + a^2\,d\,f\,z_3 + 
       a\,b^2\,f\,z_4 \right)  \right)  - b^5\,f\,z_3\,z_5 = 0 
\label{modulit2}
\end{eqnarray}
where we have relabeled the mesons $n_4$, $q_1$, $n_8$, $m_6$, $n_1$,
$n_5$ respectively with the complex variables: $z_4$, $z_3$, $z_1$,
$t$, $z_2$, $z_5$, as in the toric case (\ref{name}). Once linear
relations among mesons are kept into consideration, the quartic
relation $m_5 m_6 n_7 n_8=m_7 m_8 n_5 n_6$ can be deduced from
equations (\ref{modulit2}). 
We see therefore that the moduli space of the $T^2$ theory is a non
complete intersection in $\mathbb C^6$ of these 5 quadrics; it is easy
to verify that the dimension of the complex cone is three. 

Indeed not all the parameters appearing in the equations
(\ref{modulit2}) are associated with independent complex deformations:
to count how many complex deformations of the non toric $T^2$ cone are
admitted (that leave the manifold a cone) we have to consider how many
parameters among the $(a,b,d,f)$ can be reabsorbed
through generic linear redefinitions of the complex variables $z_1, \ldots 
z_5,\, t$. It is easy to see that through suitable rescalings of the $z_1,
\ldots z_5,\, t$ only three parameters can be reabsorbed \footnote{Note that using
  only rescalings of the chiral fields one can reabsorb only two
  parameters. Therefore differently from the toric case here it is not enough
  to consider only rescalings of the chiral fields.}. We also explicitly
checked that the remaining parameter cannot be reabsorbed through generic
linear redefinitions of the complex variables. Hence the non toric $T^2$ cone
(with the choice of relations in (\ref{3dcond})) 
admits one complex parameter of deformations. 

The case of the $(P)dP_4$ theory with isometry $T^1$ is analogous to
the case $T^2$ studied in this Appendix, even though the computations are much
more longer. We checked that for suitable values of the
coefficients in the superpotential (\ref{wt1}) the moduli space can
still be expressed as a non complete intersection of 5 quadrics in
$\mathbb C^6$. As explained in subsection \ref{t1t2ex}, typically we expect no
complex deformation for the $T^1$ cone that leaves it a cone.

\end{document}